\documentclass[twocolumn]{aastex61}

\usepackage{comment}
\usepackage{multirow}
\usepackage{amsmath}

\accepted{for publication in ApJ on Jan 18 2018}

\shorttitle{A non universal IMF. Optical data of 6 UFDs}
\shortauthors{Gennaro et al.}

\begin{document}

\title{Evidence of a non universal stellar Initial Mass Function. \\Insights from HST optical imaging of 6 Ultra Faint Dwarf Milky Way Satellites\footnote{Based on observations made with the NASA/ESA {\it Hubble
Space Telescope}, obtained at the Space Telescope Science Institute, which
is operated by the Association of Universities for Research in Astronomy, 
Inc., under NASA contract NAS 5-26555.  These observations are associated
with program GO-12549.
}}

\correspondingauthor{Mario Gennaro}
\email{gennaro@stsci.edu}

\author[0000-0002-5581-2896]{Mario Gennaro}
\affiliation{Space Telescope Science Institute,
3700 San Martin Drive,
Baltimore, MD 21218, USA}

\author{Kirill Tchernyshyov}
\affiliation{Department of Physics and Astronomy, The Johns Hopkins University, 3400 N. Charles Street, Baltimore, MD 21218, USA}

\author{Thomas M. Brown}
\affil{Space Telescope Science Institute,
3700 San Martin Drive,
Baltimore, MD 21218, USA}

\author{Marla Geha}
\affiliation{Department of Astronomy, Yale University, New Haven, CT 06520, USA}

\author{Roberto J. Avila}
\affil{Space Telescope Science Institute,
3700 San Martin Drive,
Baltimore, MD 21218, USA}

\author{Puragra Guhathakurta}
\affiliation{UCO/Lick Observatory and Department of Astronomy and 
Astrophysics, University of California, Santa Cruz, CA 95064, USA}

\author{Jason S. Kalirai} 
\affil{Space Telescope Science Institute,
3700 San Martin Drive,
Baltimore, MD 21218, USA}
\affil{Department of Physics and Astronomy, The Johns Hopkins University, 3400 N. Charles Street, Baltimore, MD 21218, USA}

\author{Evan N. Kirby}
\affiliation{California Institute of Technology, 1200 East
California Boulevard, MC 249-17, Pasadena, CA 91125, USA}

\author{Alvio Renzini}
\affiliation{Osservatorio Astronomico, Vicolo Dell'Osservatorio 5, 
I-35122 Padova, Italy}

\author{Joshua D. Simon}
\affiliation{Observatories of the Carnegie Institution for Science, 
813 Santa Barbara Street, Pasadena, CA 91101, USA}

\author{Jason Tumlinson}
\affil{Space Telescope Science Institute,
3700 San Martin Drive,
Baltimore, MD 21218, USA}
\affil{Department of Physics and Astronomy, The Johns Hopkins University, 3400 N. Charles Street, Baltimore, MD 21218, USA}

\author{Luis C. Vargas}
\affil{Department of Astronomy, Yale University, New Haven, CT 06520, USA}
\affiliation{SecurityScorecard Inc., 214 W 29th St, New York, NY 10001}



\begin{abstract}
Using deep HST/ACS observations, we demonstrate that the sub-solar stellar initial mass function (IMF) of 6 ultra-faint dwarf Milky Way Satellites (UFDs) is more bottom light than the IMF of the Milky Way disk. Our data have a lower mass limit of about 0.45 M$_{\odot}$, while the upper limit is $\sim 0.8$ M$_\odot$, set by the turn-off mass of these old, metal poor systems. 
If formulated as a single power law, we obtain a shallower IMF slope than the ``Salpeter'' value of $-2.3$, ranging from $-1.01$ for Leo IV, to $-1.87$ for Bo\"otes~I. The significance of such deviations depends on the galaxy and is typically 95\% or more. 
When modeled as a log-normal, the IMF fit results in a larger peak mass than in the Milky Way disk, however a Milky Way disk value for the characteristic system mass ($\sim0.22$~M$_{\odot}$) is excluded only at 68\% significance, and only for some UFDs in the sample. 
We find that the IMF slope correlates well with the galaxy mean metallicity and, to a lesser degree, with the velocity dispersion and the total mass.
The strength of the observed correlations is limited by shot noise in the number of observed stars, but future space-based missions like JWST and WFIRST will both enhance the number of dwarf Milky Way Satellites that can be studied in such detail, and the observation depth for individual galaxies.
\end{abstract}



\section{Introduction}
\label{sec:intro}

Since the pioneering work of \cite{1955ApJ...121..161S}, increasing evidence has been collected that the mass distribution of newly formed stars, the Initial Mass Function (IMF), has a constant behavior through space and time, at least in our Galaxy \citep{2010AJ....139.2679B,2002Sci...295...82K}.
Studies of the Milky Way field population and stellar clusters suggest that the vast majority of stars were drawn from a universal IMF \citep{2010ARA&A..48..339B}.

However, recently the IMF universality has been questioned by studies of extragalactic environments. 
For example, \cite{2010Natur.468..940V,2011ApJ...735L..13V,2012ApJ...760...70V} and \cite{2012ApJ...760...71C} detect signatures of a possible overabundance of low-mass stars in the spectra of giant elliptical galaxies. They interpret these signatures as an evidence of a steeper IMF at low stellar masses.
In addition, \cite{2012Natur.484..485C} measure significant variations of the IMF within a large sample of early-type galaxies. They infer this by disentangling the dark matter and stellar contributions to the galaxies' gravitational fields using integral-field stellar kinematics and dynamical modeling. The variations of the stellar mass-to-light ratios for their galaxies seem to rule out the universality of the IMF.

On the other hand, \cite{2014MNRAS.443L..69S} critically compare the results of the spectroscopic method of \cite{2010Natur.468..940V,2011ApJ...735L..13V,2012ApJ...760...70V} and \cite{2012ApJ...760...71C} and the dynamics-based one by  \cite{2012Natur.484..485C}. They find that while both methods agree in suggesting a global bottom-heaviness of the IMF over their respective observational samples, however a detailed comparison for the giant ellipticals in common between the samples shows large systematics and lack of agreement, thus casting doubts on the individual results. Furthermore, \cite{2017ApJ...845..157N} perform a similar comparison between dynamical, spectroscopic, and, in addition, lensing-based mass estimates for a sample of three nearby giant ellipticals. They show that, if the underlying IMF is parametrized as a single or broken power law, the results for the three methods disagree for 2 out of the 3 galaxies in the sample. The tension between the approaches is partly relieved when considering more flexible, including non-parameteric, IMF forms.

While the implications of a possible, even though not yet completely proven, non universality of the IMF are intriguing, they are based on indirect evidence from integrated measurements. 
As such, a complementary approach, using direct counts of resolved low-mass stellar populations is certainly necessary. 
Thanks to, and only with, the HST, such a study is possible within the system of Milky Way satellites, including the Magellanic Clouds and several smaller dwarfs.
The Milky Way satellites strongly differ from the Galaxy in a multitude of aspects (e.g., morphology, metallicity, star formation history) and therefore are the ideal targets for a study of the possible IMF variations among different types of galaxies.
Several recent studies have attempted to reach the faint end of the IMF in different Milky Way satellites.
These studies usually adopt a single power law model for the IMF, parametrized by a slope, and compare their best fit results to an average value for the Milky Way, typically $-2.3$ -- $-2.35$ \citep{1955ApJ...121..161S,2001MNRAS.322..231K}.
The IMF in our Galaxy is observed to experience a flattening, or turnover, below $\sim0.5$M$_{\odot}$ \citep{2001MNRAS.322..231K,2003PASP..115..763C}, so results of these studies need to be taken with great care when attempting a comparison with the Milky Way. The single power law parametrization is however convenient, and allows ease of comparison with, e.g., the extragalactic studies and ee will adopt it throughout this paper, together with a Log-normal parameterization.

For example, \cite{2013ApJ...763..110K}, using HST/ACS, found that the IMF in the Small Magellanic Cloud (SMC) outskirts has a slope of $-1.9^{+0.15}_{-0.10}$ (3$\sigma$ uncertainty), shallower than the typical Salpeter value of $-2.35$. Moreover, their analysis suggests that a single power law can be used to fit their observations down to the completeness limit of  $0.37 M_{\odot}$ (i.e., no turn over or flattening). However, \cite{2014prpl.conf...53O} claim that this result is not in contrast with an underlying log-normal IMF \citep{2003PASP..115..763C}, which cannot be ruled out at a $2\sigma$ level.
Recently, \cite{2013ApJ...771...29G}, also using HST/ACS, found that the sub-solar IMF slope in the ultra-faint dwarfs (UFDs) Hercules and Leo IV (slopes and $1\sigma$ uncertainties of $1.2_{-0.5}^{+0.4}$ and $1.3_{-0.8}^{+0.8}$, respectively) is much shallower than the slope for the Galaxy, 
which seems to confirm a trend in mass function slope vs. total host galaxy mass (with steeper slopes at higher galaxy mass). However, as with the SMC while the single-power law model yields very different results between the UFDs and the Milky Way, a log-normal IMF with the same parameters as for the Galaxy cannot be completely ruled out for the UFDs. 

In this paper we expand the study by \cite{2013ApJ...771...29G} by adding 4 more UFDs to the sample of Local Group galaxies with a well characterized IMF down to the low-mass stellar regime.
The 4 additional UFDs are Bo\"{o}tes I (Boo I), Canes Venatici II (CVn II), Coma Berenices (Com Ber) and Ursa Major I (UMa I).
These 4 UFDs are observed within the same HST/ACS program (HST GO Program ID: 12549, PI: T. Brown) as Hercules and Leo IV, and for consistency we include these two, previously studied UFDs, in our new analysis.
Other than just expanding the sample, we have improved our analysis by developing a new fitting technique. Our results for Hercules and Leo~IV are consistent with those of \cite{2013ApJ...771...29G}, although our refined analysis produces more robust uncertainty estimates, which are also smaller in Leo~IV.

\section{The data}
\label{sec:data}
\begin{figure*}[ht!]
\includegraphics[width=0.985\textwidth]{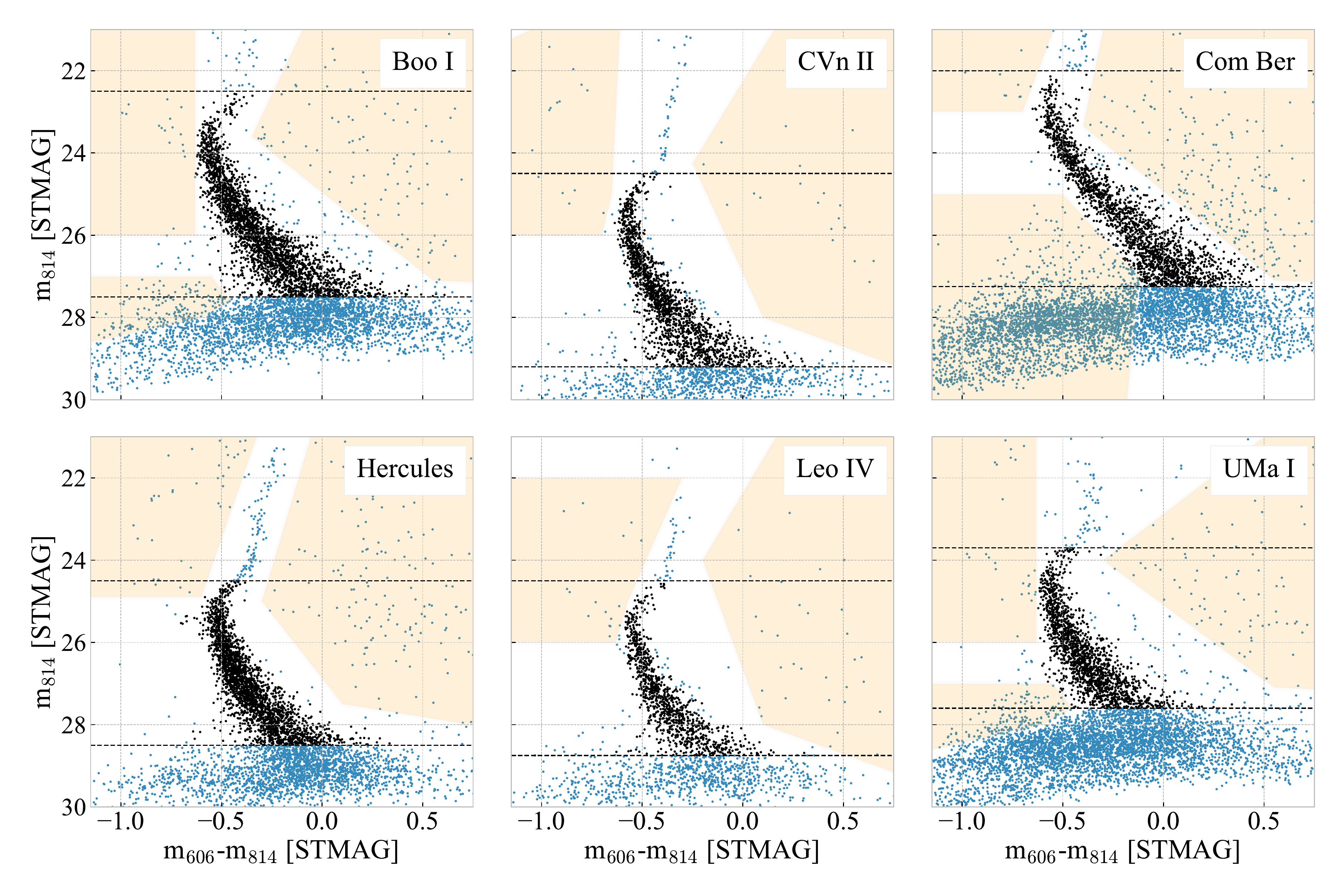}
\caption{Color-magnitude diagrams for the 6 ultra-faint dwarfs. The dashed lines represent the bright limit at the base of the red giant branch and the faint limit where both contamination and photometric errors become important. The yellow shading represents regions where no main sequence member stars are expected. The stars that are used in the analysis are in black, while those that are excluded are in blue. Stars are excluded if they are within the orange areas, above the bright limit and below the faint limit. Additionally, for the stars between the limits and outside the orange regions, a 3$\sigma$ clipping in color -- at fixed magnitude -- is used to exclude outliers.\label{fig:ALL_CMDs}}
\end{figure*}

The data used in the current analysis have already been used in \cite{2014ApJ...796...91B}. We refer the reader to that paper for a description of the observations and data reduction steps. In brief, we utilize deep optical HST/ACS images taken in the F606W and F814W filters and perform aperture (at the bright end, stars with signal-to-noise ratios $\gtrsim 100$) and PSF-fitting photometry (for all other stars).
Artificial star tests are also performed, to recover the magnitude-error distributions. These test are used in the rest of the paper for synthetic color-magnitude diagram (CMD) realizations.
With respect to the \cite{2014ApJ...796...91B} paper, we further select the data to work with a sample that excludes most contaminants (background galaxies and field stars). 
The CMDs of all 6 UFDs are shown in Fig.~\ref{fig:ALL_CMDs}.

The selection process consists of several steps: 
\begin{enumerate}
\item We first identify regions of the CMD that are obviously far from the galaxy's main sequence; such regions are visible in the figure. Stars in these regions are excluded.
\item We identify the base of the red giant branch (RGB) and exclude stars brighter than that point. The reason for doing this is to avoid problems with small color mismatches between the models and data. Such mismatches would systematically affect the fitting procedure. Excluding RGB stars has no impact on the IMF fitting for two reasons: first, these stars are few; second, these stars all have very similar masses, and these are very similar to the turn-off mass (not excluded), such that excluding them reduces the dynamical mass range for IMF fitting by less than a percent. The base of the RGB is identified by eye, using a comparison between observations and synthetic CMDs to identify the departure in color between the observed and predicted RGB locus.
\item We identify a faint limit below which we exclude all stars. This limit has a twofold motivation: it accounts for the increase in scatter from photometric errors and it accounts for increasing contamination on the blue side of the main sequence at fainter magnitudes. After experimenting with the Besancon galactic model\footnote{Available at \url{http://model.obs-besancon.fr/}} \citep{2003A&A...409..523R} we can attribute this contamination to Galactic halo white dwarfs. Given that we stop considering stars well above where the ``plume'' of contaminants approaches the galaxies' main sequences (in color), we do not attempt to better model such contaminants. We prefer to lose some of the faint UFD stars as the purity of the sample is deemed here more important than its completeness. At the same time, the data 50\% completeness limit is typically within 0.1 magnitudes of the adopted faint limit, and with a mass-magnitude relation of 0.1 M$_\odot$/mag at 0.5 M$_{\odot}$ (approximately the mass limit for all the 6 UFDs), the corresponding loss in mass range depth is less than 0.01 M$_{\odot}$, thus negligible.
\item After applying all the above selection criteria, the data are binned in 0.1 mag wide bins in m$_{814}$ and a clipping in  m$_{606}$-m$_{814}$ color is applied\textbf{, using the \texttt{scipy.stats.sigmaclip} module. The mean and standard deviation, $\sigma$, of the color of the stars in each bin is computed and stars are clipped if their colors differ more than $3.5 \, \sigma$ from the mean. The process is iterated until no more stars fall outside the $3.5 \, \sigma$ interval}. The stars excluded this way appear as blue dots outside the shaded area in Fig.~\ref{fig:ALL_CMDs}.
\item The same criteria are applied to the artificial star tests.

\end{enumerate}

Table~\ref{tab:detnumbers} reports the total numbers of selected stars and the brightness limits we imposed in the selection.


\begin{deluxetable}{lccc}
\tablecaption{Number of selected stars for each Ultra Faint Dwarf galaxy \label{tab:detnumbers}}
\tablehead{
\colhead{Galaxy} & \colhead{Selected stars} & \colhead{Bright limit} & \colhead{Faint limit} \\
\colhead{} & \colhead{[number]} & \colhead{[m$_{814}$]} & \colhead{[m$_{814}$]} 
}
\startdata
Boo I & 2698 & 22.5 & 27.5 \\
CVn II & 1727 & 24.5 & 29.2 \\
Com Ber & 2011 & 22.0 & 27.25 \\
Hercules & 2457 & 24.5 & 28.5 \\
Leo IV & 1016 & 24.5 & 28.75 \\
UMa I & 1678 & 23.7 & 27.6 \\
\enddata
\end{deluxetable}

\section{The Stellar Models}
\label{sec:models}

We adopt the same $\alpha$-enhanced stellar models as in \cite{2014ApJ...796...91B} computed using the Victoria-Regina evolutionary code \citep[see][and references therein]{2014ApJ...794...72V}, in the metallicity range $-4.0 <$ [Fe/H] $<-1.0$ dex. These models have [$\alpha$/Fe] = +0.4 dex, and a further oxygen enhancement, $\Delta$[O/Fe], that increases at decreasing [Fe/H] values \citep[details in Section 3.2 of][]{2014ApJ...796...91B}.
In the current work we are however very close to the limit where such models with enhanced abundances have been calculated, namely 0.4~M$_\odot$. Having models for masses below 0.4~M $_\odot$ is important for the simulation of binary stars that have a component less massive than this limit. At the same time, due to photometry errors, some stars with M~$<0.4$~M$_\odot$ could be up-scattered above the detection threshold, thus, even for single stars it is important to have models that go slightly below the average mass detection limit.
We extend the grid of models by computing $\alpha$-enhanced models without extra $\Delta$[O/Fe]. These models are then ``patched'' by forcing them to agree with the oxygen-enhanced ones at 0.4~M$_\odot$, and by assuming that a rigid shift applies to the model for all masses below 0.4~M$_\odot$. This is not a perfect solution, but the shift amounts to at most few hundredths of magnitudes in a region of the CMD affected by a much larger photometric scatter.  Indeed, this solution is only an approximation for those stars that are scattering from the edge of the detection limit into our fitting region.

\section{Fitting Technique}
\label{sec:method}
The fitting technique used in this work is based on proposing possible stellar population parameter sets, simulating individual CMDs according to those parameter sets, comparing the simulated CMDs to the observed CMD, and keeping or discarding the proposed stellar population parameter sets according to the result of the comparisons.
These comparisons are done in a way that ensures that the distribution of the kept parameter sets is close to the probability distribution over stellar population parameters defined in \citet{2015ApJ...808...45G}.
We use this new technique instead of the one described in the above work or CMD template fitting \citep[see e.g.][]{Harris:2001fj,Vergely:2002ve,Dolphin:2002lr,Ng:2002gf,Cignoni:2006vn,Aparicio:2009uq} because for our problem, repeatedly simulating individual CMDs is less computationally intensive than computing the tables of likelihood functions required for the \citet{2015ApJ...808...45G} approach or the CMD templates required for the template fitting approach.

The mechanics of the fitting procedure can be broken into two largely independent parts: a function for quantitatively comparing CMDs and an algorithm for using that comparison to explore a probability distribution.
An observation of one star or unresolved stellar system is a point in $\mathbb{R}^d$, where $\mathbb{R}$ is the set of real numbers, and $d$ is the number of bands used in the CMD.
Comparing CMDs requires a definition of a distance between unordered collections of points in $\mathbb{R}^d$.
There are multiple ways of defining such a distance, each with its own advantages and disadvantages.
We use the so-called \emph{kernel distance} \citep{DBLP:journals/corr/abs-1103-1625}.
The kernel distance can be computed in closed form and fulfils the assumptions of some useful theorems on the convergence of our algorithm for interpreting comparisons between CMDs.
We will describe the kernel distance in more detail in Sect.~\ref{sec:KD}.

The algorithm we use to interpret comparisons between simulated CMDs and the observed CMD is called Approximate Bayesian Computation Markov Chain Monte Carlo (ABC-MCMC, \citealt{2003PNAS..10015324M}).
ABC-MCMC is a specific form of Approximate Bayesian Computation, or ABC.
ABC is an approach to Bayesian inference designed for problems where evaluating the actual posterior probability distribution is computationally infeasible or actually impossible.
While astronomers have invented many ABC-like techniques, the conscious use of ABC in astronomy started with \citet{2012MNRAS.425...44C} and \citet{2013ApJ...764..116W}.
Knowing that a technique is an example of ABC is useful because there is a large technical literature on the statistical properties of many types of ABC.
For example, in Sect.~\ref{sec:KD} we will use a theorem from the ABC literature to affirm our choice of comparison function.
We will describe ABC in more detail in Sect.~\ref{sec:ABC}.

Appendix \ref{sec:validation} shows a series of tests we have performed to empirically validate our fitting technique.
In these tests, we simulate catalogs that are meant to resemble our Hercules galaxy dataset and attempt to recover the input stellar population parameters.
The tests demonstrate that our technique has the ability to accurately recover IMF parameters for both functional forms used in this paper (single power law IMF and log-normal IMF).
The recovered IMF parameters are, on average, unbiased even though we are only working with a very limited mass range of 0.5 -- 0.8 M$_\odot$.
The limited mass range does, however, mean that the variance of our IMF parameter estimates is fairly high; it also means that we cannot select between the two possible functional forms.

\subsection{ABC and ABC-MCMC}
\label{sec:ABC}

ABC is a technique for doing inference when generating realizations of a model given input parameters is simple but evaluating the model likelihood is inconvenient or impossible.
In the CMD fitting problem, we are dealing with the inhomogenous Poisson point process (PPP) model of a resolved stellar population, which was described in \citet{2015ApJ...808...45G}.
In that work, the likelihood of an observed flux given an intrinsic flux was assumed to be a simple, closed-form function.
The assumption of a closed-form likelihood allowed us to numerically compute the integrals involved in the PPP likelihood function (see, e.g., Eqs.~11 -- 14 of \citealt{2015ApJ...808...45G}) in a reasonable amount of time.
In this work, we directly use pre-computed artificial star tests instead of a closed-form function and incorporate more stellar parameters than were included in our previous work.
These two changes make our previous, exact, approach too time-consuming on our immediately available computational resources.

ABC is usually introduced via the following thought experiment from \citet{rubin1984}.
Suppose you have an observation, $\vec{y}$, and a model with parameters $\vec{\theta}$ for that observation from which you can generate simulated observations.
Randomly select a possible parameter set, $\vec{\theta}_i$, and then use the model to generate a simulated observation $\vec{y}_i$.
If the simulated observation and actual observation are exactly equal, keep $\vec{\theta}_i$; otherwise, discard it.
By repeating this procedure, one can generate draws from the probability distribution of $\vec{\theta}$ given the observations $\vec{y}$.

If $\vec{y}$ is a continuous variable, this procedure is only possible as a thought experiment -- the probability of generating an exact match to a continuous variable, such as the flux of a star, is zero.
If instead of requiring an exact match we are willing to accept a $\vec{y}_i$ which is within some distance threshold, or tolerance, $\tau$ of the actual measured $\vec{y}$, the probability of actually generating accepted $\vec{\theta}_i$ draws is no longer zero.
The cost of this computational tractability is that the $\vec{\theta}_i$ draws will now follow \emph{an approximation to} the original probability distribution.
This procedure was originally proposed in \citet{Pritchard:1999td} in the context of human population genetics.

ABC as described so far does not have guidelines on how to efficiently propose $\vec{\theta}_i$ values.
To propose $\vec{\theta}_i$ values more efficiently, we use ABC within Markov-Chain Monte Carlo \citep[MCMC, see e.g.][]{2003PNAS..10015324M}.
ABC can be dropped into an existing MCMC approach by treating the acceptance or rejection of a proposed value within ABC as a noisy estimate of the likelihood function.
When deciding whether to accept or reject an MCMC step, the likelihood term in the numerator of the MCMC acceptance ratio is 1 if the proposed $\vec{y}_i$ is close enough to $\vec{y}$ and 0 if it is not.
The likelihood term in the denominator is 1 by definition, since the ABC likelihood of an accepted value is 1.
The rest of the MCMC acceptance ratio is unchanged.

We adopt the \cite{2013PASP..125..306F} python implementation of the \cite{2010CAMCS...5...65G} affine-invariant sampler.
We start our MCMC runs with large values of the tolerance, $\tau$, and typically run 80 chains for 30 steps. We then use the 90\% quantile of the simulated-observed CMD distance values across the 80 walkers to define a new tolerance level. This means that at the beginning of a new iteration of 30 runs, 10\% of the walkers will find themselves outside of the acceptable parameters zone. They will be slowly drawn in by mixing with the now 90\% of good walkers. As soon as all of the walkers are in an acceptable state, the iteration is finished, a new tolerance computed and a new iteration begins.
We continue to adapt the tolerance until the new value differs from the old by less than 0.2\%. At this point we run the chains for 2500 steps.
We further throw away the first 500 steps and thin the chains by a factor of 50, a value that seemed appropriate for our MCMC experiments once the results were obtained and the autocorrelation times estimated. This way we are left with independent draws from a good approximation of the true posterior.

\subsection{Kernel Distances}
\label{sec:KD}
In order to decide whether to accept or reject a proposed parameter set in the ABC procedure described above, we need to compute the distance between a simulated dataset and the actual dataset; in the terms of our problem, we need to compute a distance between a simulated and observed CMD in order to decide whether to accept a stellar population parameter set.
We have chosen the kernel distance for this purpose \citep{DBLP:journals/corr/abs-1103-1625}.
If $\mathcal{X}$ and $\mathcal{Y}$ are collections of points in $\mathbb{R}^d$ and $K(\vec{x}, \vec{y})$ is a \emph{kernel function}, i.e. a function which maps pairs of inputs to the interval $[0, 1]$, then the square of the kernel distance between $\mathcal{X}$ and $\mathcal{Y}$ is
\begin{equation}
\begin{split}
\rho^2_K(\mathcal{X},\mathcal{Y}) &\equiv  \sum_{\vec{x}\in \mathcal{X}}\sum_{\vec{x}'\in \mathcal{X}}K(\vec{x},\vec{x}\,') + \sum_{\vec{y}\in \mathcal{Y}}\sum_{\vec{y}'\in \mathcal{Y}}K(\vec{y},\vec{y}\,') \\ & - 2\sum_{\vec{x}\in \mathcal{X}}\sum_{\vec{y}\in \mathcal{Y}}K(\vec{x},\vec{y}).
\end{split}
\end{equation}
If the kernel function is positive definite, the corresponding kernel distance is a metric, meaning that it is non-negative, symmetric, obeys the triangle inequality, and evaluates to 0 if and only if $\mathcal{X}$ is equal to $\mathcal{Y}$.
We have chosen the Gaussian kernel, which is positive definite.

In our implementation the choice of sigma depends on the data.
We first standardize our CMD into the $(0,1)\times(0,1)$ rectangle.
Working in this system ensures that similar weight is given to the magnitude and the color dimension, which are naturally quite inhomogeneous (the magnitude range of our observations being about 10 times larger than the color range).
We then compute the average minimum distance between the points (i.e., the average of the minimum euclidean distance between each point and all other points).
In the case of a uniform distribution this number is approximately $1/N^2$, but our data are not uniform.
We then set sigma to 3 times this value.
The idea is that $\sigma$ is chosen so that within each kernel there will be a few neighbors.
If $\sigma$ is too big, the kernel distance essentially returns the square root of the square difference in the number of simulated and observed stars; if $\sigma$ is too small, it returns the square root of the square sum of the number of simulated and observed stars that do not exactly line up.

While we are not aware of any prior examples of ABC being done using the kernel distance, \citet{2017arXiv170105146B} have performed an empirical and theoretical study of ABC with the Wasserstein, or Earth mover's, distance.
The Wasserstein distance is a metric on collections of points that has good theoretical properties but is non-trivial to actually compute.
\citet{2017arXiv170105146B} establish conditions under which the Wasserstein distance-ABC probability distribution converges to the actual probability distribution as the tolerance $\tau$ is reduced (Prop.~5.1).
The kernel distance meets the two assumptions required by that Proposition, meaning that the kernel distance-ABC probability distribution also converges to the actual probability distribution.
We have not determined whether the other properties of Wasserstein distance-ABC also carry over to kernel distance-ABC; this is left to future work.

\subsection{IMF Parametrization}
\label{sec:IMFpar}
We assume two possible forms for the IMF: a single power law, and a log-normal distribution. 
Our parametrization describes the system mass function, not the individual stars' mass function.
We only consider stellar systems made of one or two stars -- no higher order multiples are included.
We thus have to specify a model for binarity. This model is very simple: one parameter is used to characterize the binary fraction, $bf = \frac{\# \mathrm{binaries}}{\# \mathrm{singles} + \# \mathrm{binaries}}$. We also define a fixed distribution (i.e., one we do not fit) for the binary mass ratio, i.e., $q = \frac{m_s}{m_p}$, assuming a uniform distribution for q between 0 and 1.

Another unknown in the problem is the number of formed stars, which correspond to the intensity of the underlying inhomogeneous Poisson Point Process \cite{2015ApJ...808...45G}.
The intensity is the mean number of formed stars, with the actual number of formed stars being drawn from a Poisson distribution with mean equal to the intensity. The intensity can be seen as an overall, unknown, normalization and thus depends on the range of masses over which the IMF model is specified; in our case this range is $(0.35, 8)$~M$_{\odot}$.

We thus have 3 parameters to fit in the case of the single power-law (SPL) model: intensity, power law slope, and binary fraction; for the log-normal (LN) model there are 4 parameters: intensity, binary fraction, $m_c$, and $\sigma$. The mass distributions are specified as
\begin{eqnarray}
p_{\mathrm{SPL}}(m|\alpha) & \propto & m^{\alpha} \\
p_{\mathrm{LN}}(m|m_c, \sigma) & \propto & \frac{1}{m} e^{-\frac{1}{2}\left(\frac{\log m-\log(m_c)}{\sigma}\right)^2 \label{eq:ln}}
\end{eqnarray}
Note that the log-normal formulation above is non-standard, due to the use of the base-10 logarithm (indicated by $\log$), instead of the base-$e$ one (indicated by $\ln$). We keep this notation in order to allow for a direct comparison with the system mass function of \cite{2003PASP..115..763C}, where  $m_c$ is 0.22 M$_{\odot}$, and $\sigma$ is 0.57 (see their equation 18).
To obtain a standard log-normal, the base-10 logarithms in equation~(\ref{eq:ln}) can be replaced by base-$e$ ones, while $\sigma \rightarrow \sigma' = \sigma \times \ln(10)$.

\subsection{Assumptions In Generating Synthetic CMDs}
\label{sec:priors}
The position of a stellar system in the CMD does not only depend on its mass and binary properties, but also on its age, metallicity, distance, and extinction in each band. Rather than simultaneously fit for all these properties, we adopt priors on their distribution.

We use the star formation histories (SFHs) derived for the 6 UFDs in \cite{2014ApJ...796...91B} as priors on age for our simulations.
The SFHs are described as a 2-burst model, with each burst being a normal distribution with  $\sigma = 0.1$ Gyr, and mean value given by Table~2 of \cite{2014ApJ...796...91B}.
All the SFHs are dominated by one of the 2 normal distributions, except for UMaI for which we have 2 almost identical bursts of star formation at 14.1 and 11.6 Gyr. The mean ages are in the interval 12.7 -- 13.9 Gyr over the sample of UFDs.

Similarly to the SFHs, we adopt the metallicity distribution functions (MDFs) derived in \cite{2014ApJ...796...91B}, based on spectroscopic measurements of [Fe/H]. \cite{2014ApJ...796...91B} did not assume a functional form for the MDFs, but performed a non parametric fit to tabulated MDFs sampled at 0.2 dex interval in the [-4.0,-1.0] dex range \citep[see the shaded areas in Fig.~3 of][]{2014ApJ...796...91B}.

We also adopt the same $A_V$ values and apparent distance moduli from that paper, together with the assumption that the reddening law can be described as in \cite{1989ApJ...345..245C}, to derive values of the true distance modulus, and of the extinction in each band, $A_{606}$ and $A_{814}$. We note that \cite{2014ApJ...796...91B} assume the extinction law by \cite{1999PASP..111...63F}; this law and the one by \cite{1989ApJ...345..245C} are nearly identical in the wavelength regime bracketed by our photometric bands. 
Given their distance it is in practice reasonable to assume that all of the stars in each of the 6 UFDs are at the same distance, and thus we adopt a single value for the prior on distance. We also assume that all stars are subject to the same reddening, so fixed global values for each UFD are used as priors on the individual photometric bands' extinction.

\subsection{Summary Of The Method}
For each ABC-MCMC step we have a tuple of parameters describing the underlying Poisson Point Process: the intensity, $\Lambda_i$, one or two parameters describing the IMF distribution, $\vec{\theta}_{\mathrm{IMF},i}$, and one parameter for the binary fraction, $\vec{\theta}_{\mathrm{Bin},i}$. We extract an integer number, $n_i$, of ``born'' stellar systems from a Poisson distribution with mean equal to $\Lambda_i$. We extract $n_i$ values of masses and binary properties from $p(\vec{\theta}_{\mathrm{IMF},i})$ and $p(\vec{\theta}_{\mathrm{Bin},i})$, respectively. We also extract $n_i$ ages, metallicities, extinction values in each band and distances from the immutable prior distributions of Sect.~\ref{sec:priors}.

We then compute the corresponding magnitudes given mass, age, binary fraction, metallicity, extinction and distance, using the stellar models and the artificial stars experiments. Some of the simulated objects will not be observable due to either stellar evolution or incompleteness. We thus end up with $m_i \leq n_i$ stars in the CMD.

We compute the kernel distance between the $m_i$ simulated points and the actual observations (see Sect.~\ref{sec:KD}), and accept/reject the proposed ($\Lambda_i, \, \vec{\theta}_{\mathrm{IMF},i}, \, \vec{\theta}_{\mathrm{Bin},i}$) tuple according to the ABC method described in Sect.~\ref{sec:ABC}. We finally iterate using the ABC-MCMC scheme described in Sect.~\ref{sec:ABC} as well.

\section{Results}

\begin{deluxetable*}{l|cc|cc|cc||cc|cc|cc|cc}[t]
\tablecaption{Best fit parameters for both the adopted IMF models\label{tab:res}. Unif and log column labels stand for parameter estimations with a uniform prior in Intensity or a prior that is uniform in $\log(\mathrm{Intensity})$, respectively. Full versions of the results, including also the credible intervals are given in \textbf{Appendix~\ref{sec:fulltabs}}, Tables ~\ref{tab:res_long_SPL} and ~\ref{tab:res_long_LN} for the single power law and log-normal IMF models respectively.}
\tabletypesize{\footnotesize}
\tablehead{
\multirow{3}{*}{Galaxy} & \multicolumn{6}{c||}{Single Power Law} & \multicolumn{8}{c}{Log-normal} \\
 &  \multicolumn{2}{c}{Intensity} & \multicolumn{2}{c}{Slope} & \multicolumn{2}{c||}{Bin. Frac} & 
    \multicolumn{2}{c}{Intensity} & \multicolumn{2}{c}{m$_c$[M$_{\odot}$]} & \multicolumn{2}{c}{$\sigma$} & \multicolumn{2}{c}{Bin. Frac}  \\
       & unif & log & unif & log & unif & log & unif & log & unif & log & unif & log & unif & log
}
\startdata
Boo I      & 8.39e+3 &  8.28e+3 &   -1.84  &  -1.87  &  0.28  &  0.28& 7.75e+3 &  7.46e+3 &  0.33 & 0.34 & 0.59 & 0.56 &  0.27 & 0.28 \\ 
CVn II     & 1.58e+4 &  1.36e+4 &   -1.00  &  -1.17  &  0.05  &  0.06& 9.91e+3 &  9.51e+3 &  0.56 & 0.54 & 0.70 & 0.67 &  0.06 & 0.07 \\ 
Com Ber    & 3.71e+3 &  3.67e+3 &   -1.66  &  -1.68  &  0.61  &  0.61& 3.52e+3 &  3.45e+3 &  0.30 & 0.30 & 0.67 & 0.65 &  0.60 & 0.60 \\ 
Hercules   & 2.75e+4 &  2.32e+4 &   -0.93  &  -1.11  &  0.12  &  0.13& 1.64e+4 &  1.60e+4 &  0.60 & 0.59 & 0.69 & 0.67 &  0.11 & 0.11 \\ 
Leo IV     & 1.20e+4 &  1.01e+4 &   -0.82  &  -1.01  &  0.13  &  0.14& 6.40e+3 &  6.14e+3 &  0.62 & 0.61 & 0.67 & 0.64 &  0.14 & 0.14 \\ 
UMa I      & 1.05e+4 &  9.56e+3 &   -1.46  &  -1.58  &  0.30  &  0.31& 8.19e+3 &  7.92e+3 &  0.53 & 0.51 & 0.64 & 0.61 &  0.30 & 0.31 \\ 
\hline
\hline
\enddata
\end{deluxetable*}

We run our ABC-MCMC experiments adopting uniform priors on the IMF model parameters. It is however interesting to note that, for the single power law case, there is a strong correlation between the intensity and the power law slope, especially at shallower slopes. 
Triangle plots showing the pairwise relations between the parameters are shown in Appendix~\ref{sec:corner_plots} \citep[such plots make use of the \texttt{corner.py} python code,][]{corner}.

The correlation between slope and intensity is due to the fact that for a higher total number of stars but a shallower (less negative) slope, the number of observed stars remains almost constant. 
Given the narrow mass range probed, also the shape of the luminosity function (and the CMD) shows only very minor changes. It is however well established that galaxies with less stars are more probable than larger galaxies. Thus we explored the idea of assuming a prior on the total number that is flat in logarithm, rather than in actual number.
We do not run separate MCMC experiments, but we rather weight the results assigning to each draw a weight that is inversely proportional to the value of the intensity for that draw. This is equivalent to assuming a logarithmic prior in the intensity.
For completeness we report results obtained both ways. We note that in some cases, as for Boo~I  or Com~Ber, the changes are minor, thus the prior seems to be of little importance here. However, for other galaxies like CVn~II, Hercules and Leo~IV, the difference in the average slope is of the order of 0.2, which is not negligible.
In general the prior, by downweighting the highest intensities, correspondingly downweights the shallowest slopes, thus in all cases a steeper slope is reported when using the more realistic logarithmic prior.

The results are summarized in Table~\ref{tab:res}, which reports the 
best-fit values for the IMF models
parameters. A visualization of the results is given also in Figure~\ref{fig:marg}.
A more exhaustive summary of the results is given in appendix~\ref{sec:fulltabs}, where Table~\ref{tab:res_long_SPL} and Table~\ref{tab:res_long_LN} contain not only the best-fit values but also the 68\%, 95\%, and 99\% credible intervals (CIs) for the parameters.
Figure~\ref{fig:simfrombf} shows instead a comparison of the observed $m_{814}$ luminosity functions and of the luminosity function that can be obtained by using the best-fit parameters to simulate a CMD. This figure also shows the histogram of the masses of the simulated stars, for both those that land in the CMD (the present day mass function), and for all the extracted ones, including those that are too faint to be observed and those that have evolved off the main sequence (the initial mass function).

In reference to the mass histrograms of figure~\ref{fig:simfrombf}, we stress that even though the observed mass range for single stars is limited to about 0.5--0.8 M$_{\odot}$, the IMF model for stellar systems needs to be specified on a wider mass range, 0.35--8.0 M$_{\odot}$, see Section~\ref{sec:IMFpar}. This extension ensures that the secondary stars from more massive systems as well low-mass single stars with large photometric errors that are up-scattered in the CMD are correctly accounted for. We also enforce that the model range specification is the same for all UFDs. The implication of our assumptions is obviously that if the IMF functional form is very different outside the specified range, then the model parameters will be systematically affected. On the other end, a partial model specification, limited to a very narrow mass range, would equally bias the results by neglecting binary secondaries and low mass objects. 

\subsection{Single Power Law Model}
\begin{figure*}[ht!]
\includegraphics[width=0.985\textwidth]{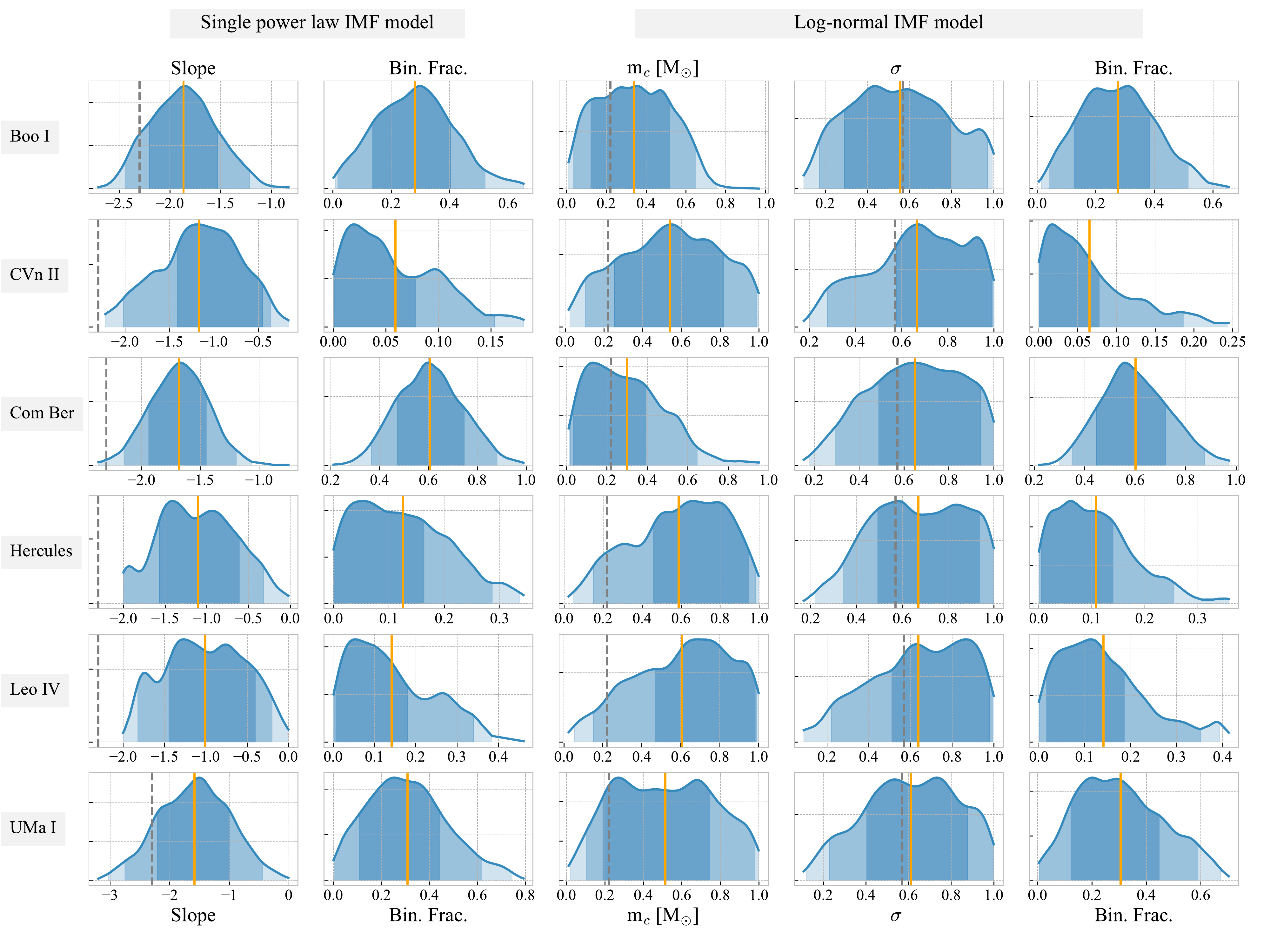}
\caption{Gaussian Kernel Density estimates of the marginal probability distributions for the fitted parameter for all the UFDs. The kernel width, $w$, is estimated using Scott's rule \citep{2015mdet.book.....S}: $w = \sigma  n^{-1./(d+4)}$, with $\sigma$ being the data variance, $n$ the number of points, $d$ the number of dimensions (always 1 in this case).  The orange lines correspond to the best-fit values, i.e., the average of the marginals. The shaded areas indicate credibility intervals, estimated as the smallest intervals containing a $(0.6827, 0.9545, 0.9973)$ fraction of the probability. The vertical gray dashed lines correspond to the standard Milky Way Disk values, slope = -2.3 \citep{1955ApJ...121..161S}, for the single power law model, and $(m_c, \sigma) = (0.22, 0.57)$ for the Log-normal model \citep[values from the system mass function of][]{2003PASP..115..763C}. 
For this figure we use the posterior probability obtained using a flat prior on the logarithm of the intensity. \label{fig:marg}}
\end{figure*}

\begin{figure*}[ht!]
\gridline{
\fig{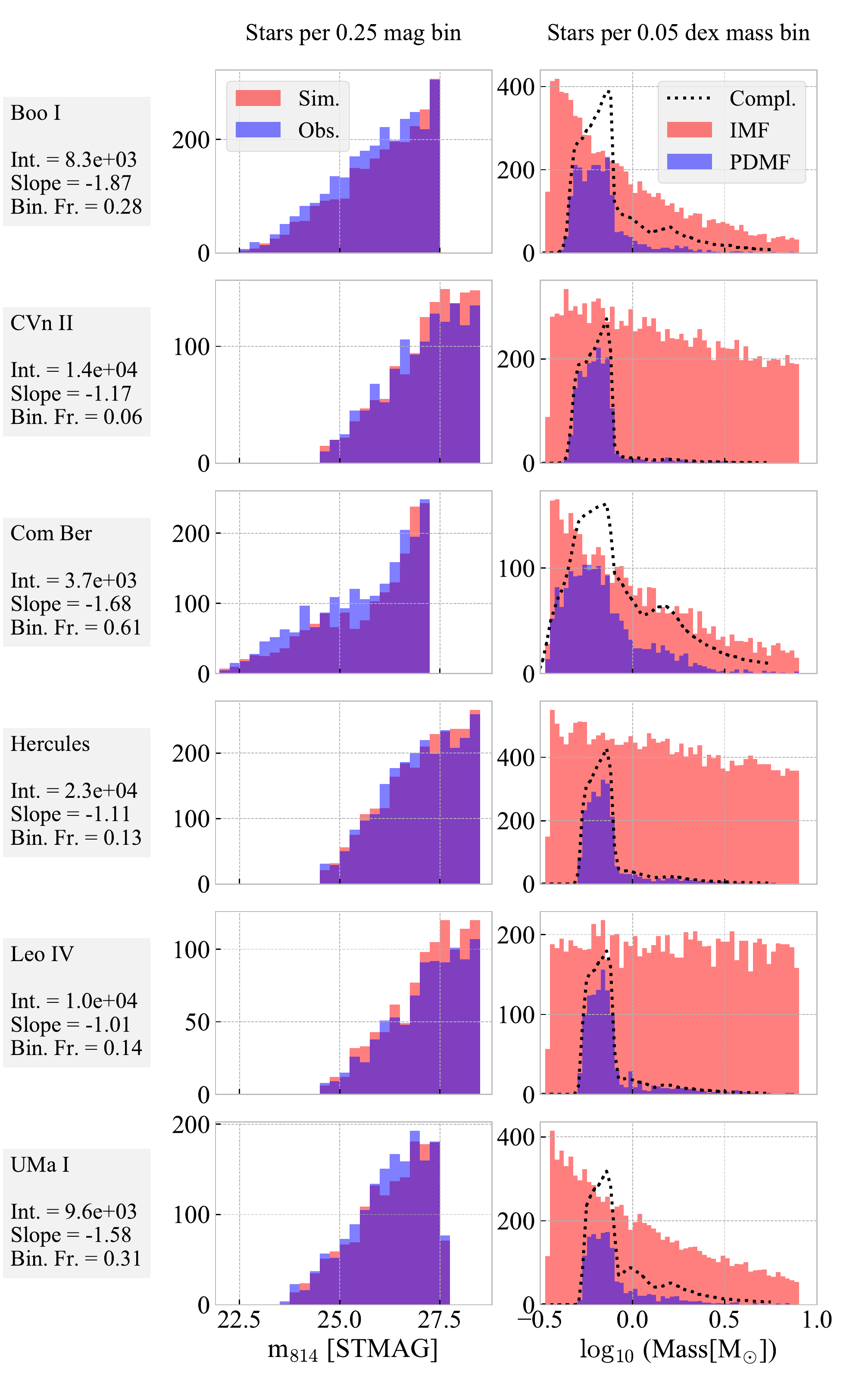}{0.485\textwidth}{Single Power Law IMF model}
\fig{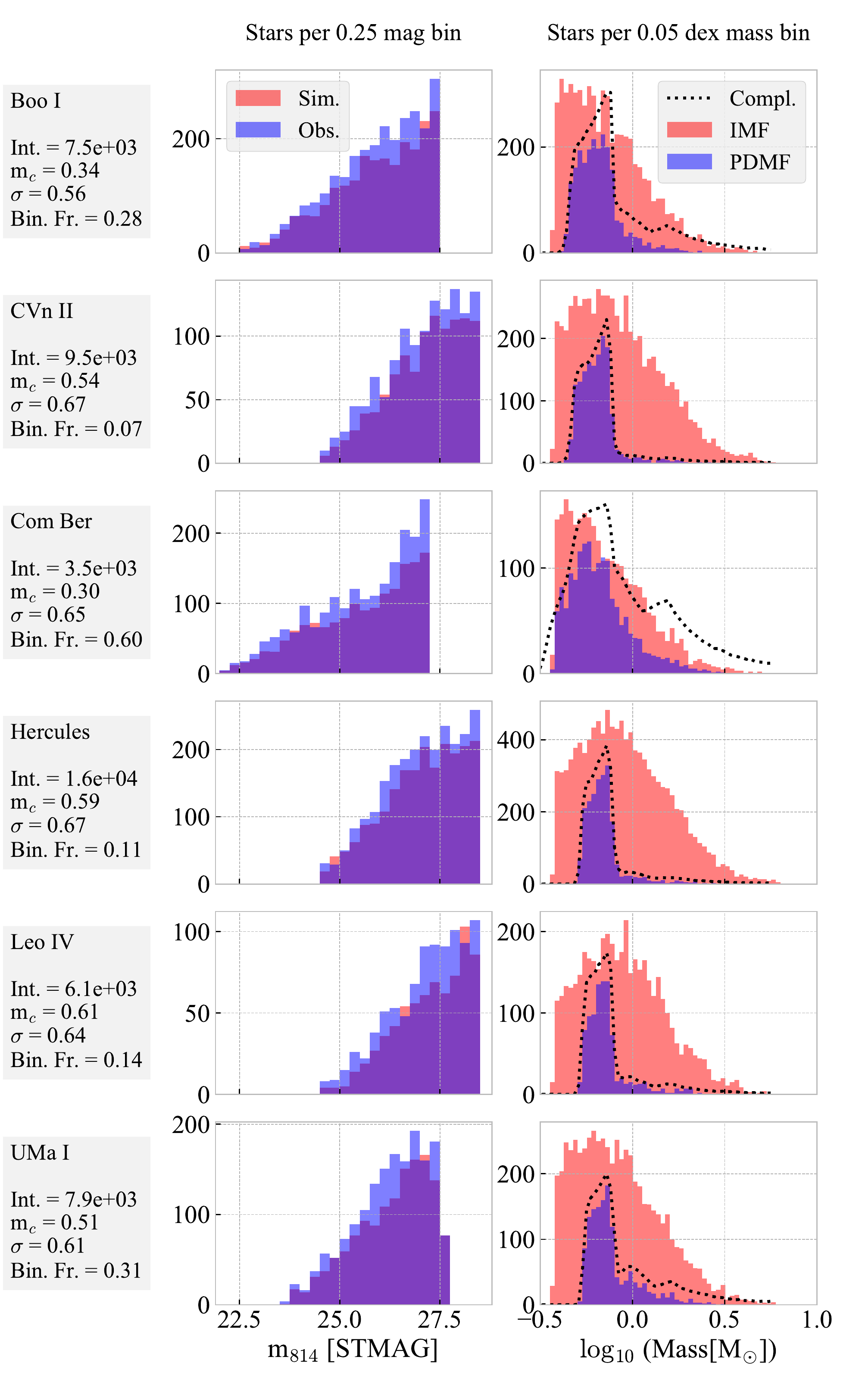}{0.485\textwidth}{log-normal IMF model}
}
\caption{Simulations obtained using the best-fit values for each UFD, for the two IMF models: single power law (left) and log-normal (right). The left columns for each IMF model show the simulated F814W luminosity function (red), compared to the observed one (blue). The right columns within each IMF model case show the full ensemble of system mass values that are drawn to obtain the luminosity function (the Initial mass function, IMF, red), as well as the masses that survive into the CMD (the present-day mass function, PDMF, blue). The dotted lines indicate the ratios between the two, i.e., the completeness as a function of system mass; the y-range of completeness values is not reported for clarity, but it goes from 0 (bottom of the y-axis) to 1 (top of the y-axis). Completeness never reaches values of 1 as a function of system mass, because there are always combinations of system mass and binary mass ratio that make it such that some systems are unobservable. The drop of completeness above $\sim 0.8 M_{\odot}$ corresponds to the turn-off mass. There are however systems that are observable even at system masses higher than the turnoff mass. These are systems where the total mass is larger than the turnoff one, but the individual components are on the main sequence and observable, thus the system is ``complete''.  \label{fig:simfrombf}}
\end{figure*}

Even if we consider the logarithmic prior case, which produces steeper slopes, for all 6 galaxies the slope values are less negative (flatter IMF) than the value for the Milky Way disk IMF, -2.35 \citep{1955ApJ...121..161S}. 
By looking at Figure~\ref{fig:marg}, and by reading the values out of Table~\ref{tab:res_long_SPL}, it is clear that for all 6 UFDs the Salpeter value falls outside the 68\% CI. For Boo~I and UMa~I the Salpeter slope is still within the 95\% CI; in the case of Com Ber, it falls outside the 95\% CI, but barely within the 99\%; while for CVn~II, Hercules, and Leo~IV the Salpeter slope is outside the 99\% CI, with the latter 2 UFDs having 99\% of their IMF slope posterior probability below -2.0. 

We note however, that a single power law model for the Milky Way IMF is not a valid description down to very low stellar masses.
\cite{2001MNRAS.322..231K} show that below 0.5 M$_{\odot}$, the IMF for the Galactic Field population I stars flattens, changing slope from -2.3 to -1.3, they thus prefer a broken power law model to describe the Milly Way IMF. Equivalently \cite{2003PASP..115..763C} parametrize this flattening as a gradual change, described by a Log-normal function, below $\sim 1$M$_{\odot}$, while they adopt the Salpeter value above this mass.
It could be argued that the low mass limit for our measurements is near the \cite{2001MNRAS.322..231K} inflection point (or within the \cite{2003PASP..115..763C} flattening region), thus our results for a single power law model could be artificially "flatter" if the underlying mass distribution for the UFDs does indeed have a similar break point.
On the other hand, studies of Milky Way star forming regions and young cluster \citep[see e.g][]{2012ApJ...748...14D,2017A&A...602A..22A}, show that for recent star formation in the Milky Way, the distribution of stellar masses can described by a single power law, with slope of $\sim -2.3$ down to 0.2--0.3 M$_{\odot}$ (it should be noted that \cite{2012ApJ...748...14D} also show that the uncertainties on pre-main sequence evolutionary models imply very large uncertainties on the derived stellar masses, and thus IMFs, in young star forming regions). 
It is still useful to compare our slopes with the traditional -2.3 value for the Milky Way, with the caveat that this single power law parametrization may not truly apply to the Galaxy throughout the full stellar mass range, with several authors slightly disagreeing about where the IMF exactly changes its behavior. 

Regardless of the limitations of a simplified single power law description, it must be noted that in several cases (CVn~II, Hercules, Leo~IV) our best fit slopes are flatter than even the Kroupa slope for M~$<0.5$~M$_{\odot}$, -1.3. Thus if we take that as the lower limit that could be expected by erroneously adopting a single power law model for a true underlying broken power law distribution, our results appear even more intriguing.
Globally these results suggest that when the IMF is described as a power law, the UFDs have a less negative IMF slope than the Milky Way, in the probed mass range (see also Sect.\ref{sec:ensemble}).
Extension of our results outside this range may however be prone to systematic errors.

\subsection{Log-normal Model}
A parallel result is that obtained for the log-normal IMF model. For all 6 UFDs, the mean characteristic system mass is found to be larger than the Milky Way disk value, $m_c = 0.22 -- 0.25 $~M$_{\odot}$ \citep{2003PASP..115..763C,2010AJ....139.2679B}, with mean values as high as $m_c \sim 0.5$M$_{\odot}$ for CVN~II, Hercules and Leo~IV.
The shallower slope for the single power law model translates into a larger characteristic mass, but the result implied by the data is the same: the UFDs have produced relatively less low-mass stars with respect to the Milky Way.
The widths of the UFDs best fit Log-normal models are in the 0.59--0.70 range, similar but also slightly larger than the Milky Way one, $\sigma = 0.57$.

The significance of the deviations of the UFDs values with respect to the Milky Way ones is however weaker for the Log-normal model than it is for the single power law one.
In more detail, the $m_c$ 68\% CIs for Boo~I, Com~Ber and Uma~I contain the \cite{2003PASP..115..763C} value of 0.22 M$_{\odot}$. For CVn~II, Hercules and Leo~IV the same characteristic mass is outside the 68\% CIs, but still within the 95\% CIs.

Given the limited range of masses we actually observe, we do not interpret this fact as the Log-normal being ``better'' at giving a more universal representation of the IMF. We rather interpret it as this model being more ``flexible'' in reproducing the data under study. Multiple combinations of $(m_c,\sigma)$ can be used to obtain similar ``slopes'' for the Log-normal distribution within the observed mass range. The flexibility added by one extra parameter, makes it such that is is harder to constrain them individually, again, due to the limited mass range.

\section{Discussion}

\subsection{Dynamical effects}
The implicit assumption in our analysis is that the distribution of stellar masses in the UFDs did not change with time. Dynamical evolution could potentially affect the IMF measurements and make our assumption invalid. For example, in globular clusters it is observed that two body relaxation, which causes mass segregation, alters the present day mass function. Lower-mass stars have higher velocities and are found at larger radii, thus escape the clusters more easily than higher-mass ones.
\cite{2012MNRAS.422.1592L} claim that when dynamical modeling is taken into account, no significant differences are measured between the IMFs of Milky Way globular clusters, and thus the stellar masses in globular clusters are consistent with being drawn from the same universal IMF. The recent study by \cite{2017MNRAS.471.3845W} does however argue that, at least for the NGC 5466 and NGC 6101 globular clusters, a non universal IMF may be required to explain the observations.

Two body relaxation times in globular clusters are of the order of 1 Gyr, thus shorter than the age of the clusters themselves, of order 10 Gyr, therefore these systems are completely collisional and strongly affected by internal dynamics.
On the other hand UFDs are systems with total masses, and thus typical internal velocities, similar to globular clusters, while stellar encounters, which are responsible for the relaxation, are much more infrequent, given the smaller number of stars.
This is a consequence of the UFDs being Dark Matter dominated.
The relaxation time, which can be defined as the time it takes for two-body encounters to change a star's velocity by an amount equal to itself, is thus much longer for UFDs than for globular clusters.

Using equation (2-62) of \cite{1987degc.book.....S}, the relaxation time can be written as:
\begin{equation}
\label{eq:Spitz}
t_{rel} [\mathrm{Gyr}] \sim \frac{3.4}{\ln(\frac{b_{max}}{b_{min}})} \left(\frac{v}{ \mathrm{km\, s}^{-1}}\right)^3 \left(\frac{m}{M_{\odot}}\right)^{-2}\left(\frac{n}{\mathrm{pc}^{-3}}\right)^{-1}.
\end{equation}
Twice the system size $2R$, can be used as maximum impact parameter, $b_{max}$.
The minimum impact parameter can be fixed as the distance for which the potential energy in an encounter equals the kinetic energy, $b_{min} = 2Gm / v^2$, i.e. the radius for which the weak encounters approximation, used to derive equation(\ref{eq:Spitz}), fails.
Typical numbers for the UFDs are $N\sim10^4$, $v\sim 4 \mathrm{km\,s^{-1}}$, $m\sim 0.6 M_{\odot}$ and $R\sim 100 \mathrm{pc}$.
The Coulomb logarithm evaluates to $\sim 13$. Substituting $n = \frac{3N}{4\pi R^3}$ we have:

\begin{equation}
t_{rel} [\mathrm{Gyr}] \sim \frac{1.1}{N} \left(\frac{v}{ \mathrm{km\, s}^{-1}}\right)^3 \left(\frac{m}{M_{\odot}}\right)^{-2} \left(\frac{R}{\mathrm{pc}}\right)^3
\end{equation}
With the above typical UFDs numbers we thus obtain $t_{rel} \sim 2\times10^4 \,\mathrm{Gyr}$, much longer than the age of the Universe. It follows that, as opposed to globular clusters, dynamics is not expected to affect the measurement of the IMF in UFDs.

\subsection{Comparison With Previous Results}
\cite{2013ApJ...771...29G} studied the IMF in Hercules and Leo IV using the same data used in this paper. Our new results confirm what was found previously. For the single power law model, the slope values agree very well within the reported $1\sigma$ uncertainties (we use the 68\% credibility interval as the equivalent measure) as can be seen in Table~\ref{tab:compG13}. The table also shows how each paper's results for the IMF slope compare to the classical values of -2.35 for M~$>0.5$~M$_{\odot}$.

For Hercules our results are nearly identical to those of \cite{2013ApJ...771...29G}.
Comparing our credibility interval to their ``sigma''-levels in a quantitative way is not straightforward. 
However, we note that our ``steep'' limit for the 99\% credibility interval is -2.0, while \cite{2013ApJ...771...29G} report that they rule out Salpeter at the 5.8$\sigma$ level. For both papers it is extremely unlikely that the masses of the stars in Hercules and in the Milky Way disk were drawn from the same distribution, independent on the adopted parametrization (similar results, although less significant are obtained for the Log-normal model as well).

For LeoIV, the reported best fit slopes are again very similar and well within the reported ($1\sigma$, or 68\% CI) uncertainties. However, our current analysis allows a more precise estimate, with uncertainties almost halved, and thus constitutes a significant improvement in terms of strength with which a Galaxy-like IMF slope can be ruled out for a single power law model. While \cite{2013ApJ...771...29G} reported that a slope of -2.3 could be only ruled out at $1.9\sigma$ level, we obtain a 99\% credibility interval steeper limit of -2.0 (the same as for Hercules).

Our analysis, even limited to just these two objects, would already strengthen the conclusions of \cite{2013ApJ...771...29G}, that ultra-faint dwarf galaxies have a different IMF than the Milky Way.

\begin{deluxetable}{l|cc|cc}
\tablecaption{Comparison of the results for Hercules and LeoIV between this work (TW) and \cite{2013ApJ...771...29G} (G13).\label{tab:compG13}}
\tabletypesize{\footnotesize}
\tablehead{
\multirow{3}{*}{Galaxy} & \multicolumn{2}{c|}{Power Law Slope} & \multicolumn{2}{c}{Comparison with } \\
                        &    &     & \multicolumn{2}{c}{\cite{1955ApJ...121..161S} }\\
                        & TW & G13 & TW & G13 
}
\startdata
Hercules   & $-1.11^{+0.50}_{-0.46}$ & $-1.2_{-0.4}^{+0.5}$ & Outside 99\% CI & $5.8\sigma$\\ 
Leo IV     & $-1.01^{+0.61}_{-0.44}$ & $-1.3^{+0.8}_{-0.8}$ & Outside 99\% CI & $1.9\sigma$\\ 
\hline
\hline
\enddata
\end{deluxetable}

\subsection{Ensemble properties}
\label{sec:ensemble}
\begin{figure*}[ht!]
\includegraphics[width=0.985\textwidth]{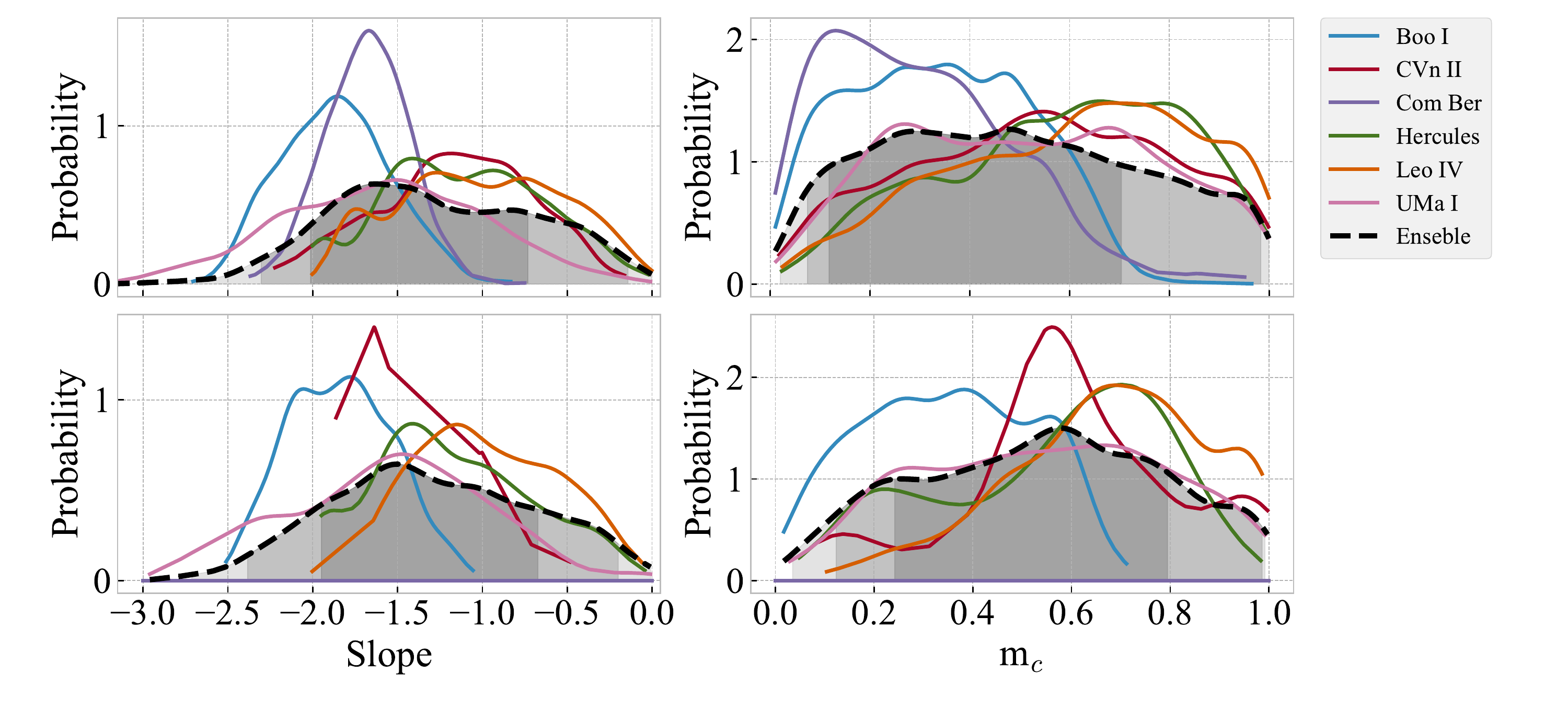}
\caption{\label{fig:ER} Left: comparison between galaxies of the IMF slope distributions for the single power law model. Right: comparison between galaxies of the IMF characteristic mass for the Log-normal model. Top: complete results. Bottom: results when only MCMC draws with a binary fraction $0.15 < \mathrm{B.F.} < 0.25$ are considered; this excludes Com~Ber from the comparison, whose MCMC solution never reaches this low binary fraction range.
In all panels the black dashed line represents the distribution over the whole UFDs ensemble, obtained by combining the individual UFDs MCMC draws. The gray shaded areas represent the credibility intervals, estimated as the smallest intervals containing a $(0.6827, 0.9545, 0.9973)$ fraction of the probability. }
\end{figure*}

All the UFDs in our sample show a flatter slope or alternatively a larger characteristic mass, when compared to the typical values found for the Milky Way disk and young cluster.
The significance of the discrepancy varies with UFD.
To evaluate the overall significance for the ensemble of UFDs, we consider the union of their MCMC draws, and look at the resulting global distributions of slope and $m_c$, shown in Fig.~\ref{fig:ER}.

The figure shows that for both slope and $m_c$, the resulting distribution over the ensemble is not much narrower than the individual distributions. For the single power law model, the value -2.3 is at the limit of the 95\% credible interval, while the $m_c$ value of 0.22 falls within the 68\% one.
IT is quite clear that the variations between galaxies are comparable to the individual uncertainties, thus combining the draws does not reduce the global uncertainty by the expected square root of six.

In reference to the same figure, it is very interesting to note that most of the scatter in the sample is caused by the differences between Boo~I and Com~Ber with respect to CVn~II, Hercules and Leo~IV, with UMa~I somewhat in between the two groups.
The common denominator between Boo~I and Com~Ber is that they are the closest UFDs in our sample with apparent distance moduli of $(m-M)_V = 19.11$ mag and 18.08 mag respectively, followed by UMa~I at 20.10 mag. CVn~II, Hercules and Leo~IV have distance moduli of  $(m -M)_V = 21.04, 20.92, 20.10$ mag instead.

These differences may suggest that some physical process, whose strength depends on the distance of these satellites from the Milky Way, may be responsible for the observed differences in IMF.
However interesting such an interpretation may be, an alternative is that the differences reside in some systematics within the data.
Boo~I, Com~Ber and UMa~I, being closer, subtend a larger angle in the sky and required more $HST$/ACS pointings to be observed (5, 12, and 9, respectively). CVn~II, Hercules and Leo~IV required only 1, 2, and 1 pointings. The wider observations translate into a larger number of fore- and background contaminants. Even though we take great care in the selection process (Sect.~\ref{sec:data}), our catalogs might be suffering from a different level of impurity for the different galaxies.
It is reasonable to expect that most of the undetected contaminants lie toward the faint end of our dataset. In this hypothesis, the effect of such interlopers would be to increase the star counts at low masses, and make the IMF of the closest UFDs steeper.

\subsection{IMF Slope and Galaxies Properties}
We investigate possible dependencies of the derived IMF slopes on several galactic properties: mean metallicity, mean age, mean individual stellar mass in the observed range, velocity dispersion, total dynamical mass, ellipticity, half-light radius and absolute magnitude.
For each of these quantities we report the Pearson correlation coefficient, $\rho$, as well as the p-value. The latter roughly indicates the probability of an uncorrelated system producing datasets that have a Pearson correlation at least as large as the one computed from these datasets. We used the python \texttt{scipy.stats.pearsonr} module for such calculations. The p-value cannot be considered completely reliable for our small sample of 6 galaxies, however we still choose to report it for completeness. 
\textbf{The p-value is used to estimate the extent to which the within-sample covariance is consistent with no-correlation. It does not however quantify the amount of correlation that might be induced by the uncertainties on the individual points. Given that these are themselves large, our correlation coefficients cannot be considered extremely robust, but a rather qualitative assessment.}

\begin{figure*}[ht!]
\includegraphics[width=0.985\textwidth]{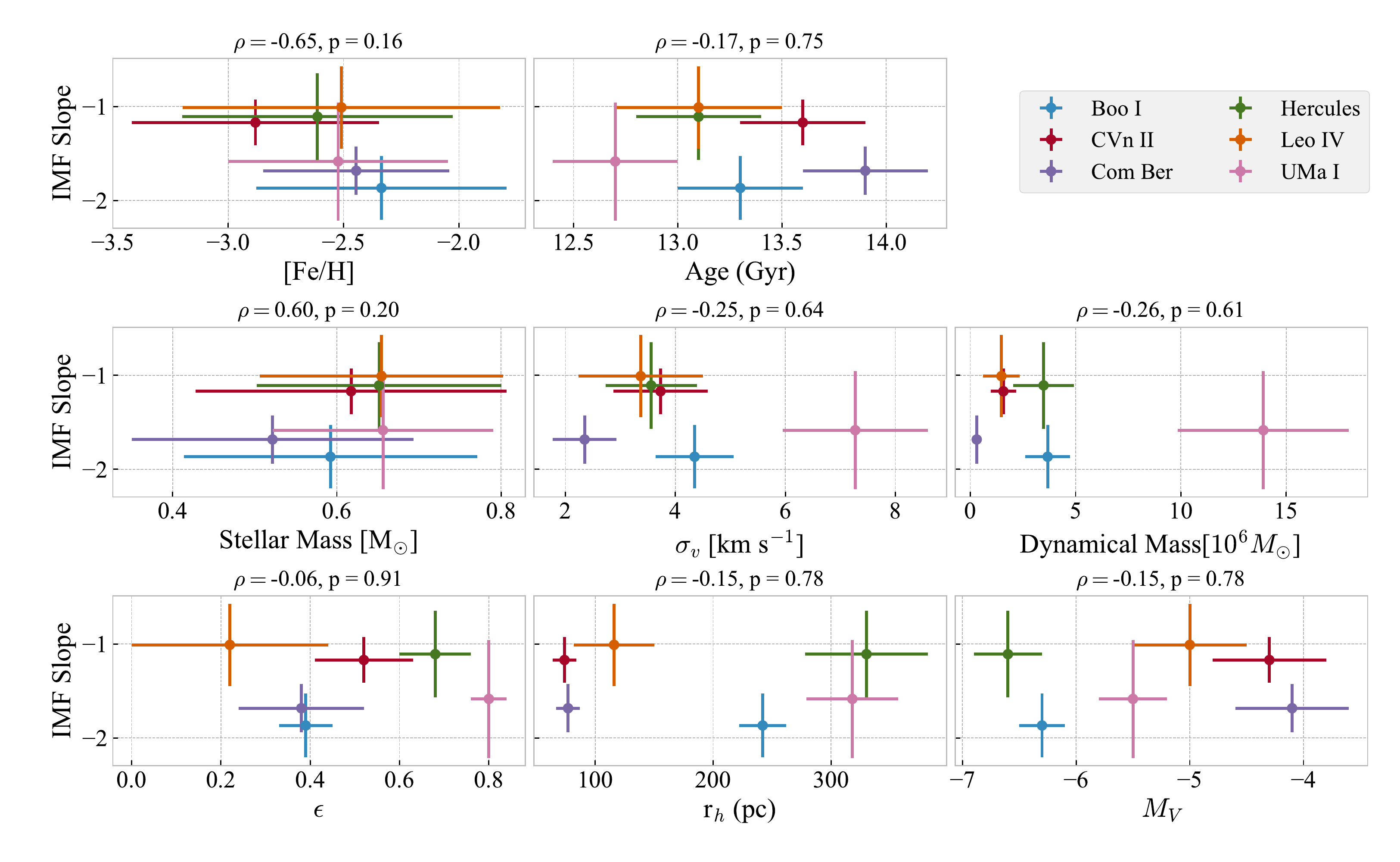}
\caption{\label{fig:trends}IMF slope trends with several galactic properties. $\rho$ is the Pearson correlation coefficient, and the p-values represents the probability of obtaining an equivalent or larger values of $\rho$ from a sample with the same number of data but no correlation. For the metallicity (top left) the range bars indicate the range of observed metallicities, not errors in the metallcities or the mean metallicity.}
\end{figure*}

\paragraph{Mean Metallicity}
The mean metallicities plotted in Figure~\ref{fig:trends} are derived from the MDFs described in Section 2.2 of \cite{2014ApJ...796...91B}. The latter are based on spectra obtained with $Keck$/DEIMOS, also described in the aforementioned paper.
The error bars on the [Fe/H] values represent the 16\% and 84\% quantiles of the MDFs, {\it not} the errors on the mean metallicities themselves.
The anti-correlation between IMF slope and mean metallicity is the strongest we observe, and also the one with the lowest p-value, i.e., the lowest chance of being a false positive. The observed correlation is suggestive that metallicity regulates the star formation process (e.g., by affecting accretion, massive stars feedback). 
\paragraph{Mean Age}
The mean ages for the stellar populations within each UFD are derived from Table 2 of \cite{2014ApJ...796...91B}. These mean values are the weighted average of the ages of a 2-burst star-formation history model, fitted to the same data used in the present paper. 
For all UFDs but UMa I, one of the age components is truly dominant thus the mean age reported here is close to one of the peaks of the bimodal age distribution model. For UMaI, the mean age is 12.7 Gyr, in between the 2 best-fit burst of 14.1 and 11.6 Gyr.
No correlation is observed between mean age and IMF slope for our sample of UFDs.

\paragraph{Mean Stellar Mass}
These mean mass values are obtained by using the best-fit parameters for the single power law model IMF and the logarithmic prior to simulate CMDs with a large number of stars. Then, only the systems that are not binaries are considered; they are used to determine the average observed mass, as well as the extrema of the observable mass range. Obviously, the mean mass value anti-correlates with the IMF slope because, by construction, steeper IMF slopes correspond to smaller average masses, provided that the mass intervals considered are similar. Even though the correlation with mean mass is not truly meaningful, we still plot the slope versus mass, because it is worth noting that the IMF slopes do not correlate with the probed mass interval (shown as the horizontal error bar). For example, the probed intervals for LeoIV and UMaI as well as for BooI and CVnII are basically the same, but the best fit slopes differ substantially.
This sanity check ensures that at least one type of systematic effect can be ruled out, i.e. the effect of an artificially changing slope that would be obtained by trying to fit a truly log-normal distribution using power law models in different intervals. 

\paragraph{Velocity Dispersion}
Velocity dispersion values are obtained from the same spectra for which we obtained the metallicity distribution functions. We use an approach similar to that of \cite{2017ApJ...838....8L}. We assume an underlying single Gaussian distribution for the ensemble of measured line-of-sight velocities and use Markov-Chain Monte Carlo methods to derive the posterior distribution of both the mean velocity and velocity dispersion, $\sigma_v$. We use the median value as the best-fit estimate, and use the 16\% and 84\% quantiles for the errors. There is some correlation of IMF slope and $\sigma_v$, but not as strong as that with metallicity. Moreover, its p-value is quite high. The $\sigma_v$  values derived here are used to derive the dynamical masses (see below). We warn the reader that further analysis is ongoing and that a more complete and thorough study of the individual velocities as well as of the galaxies' velocity dispersions are the objective of a future paper.

\paragraph{Dynamical mass}
The dynamical mass values we use are derived from our velocity dispersion values, using Eq.(3) of \cite{2007ApJ...670..313S}:
\begin{equation}
\label{eq:mtot}
\mathrm{M_{tot}} = 167\beta r_c \sigma_v^2;
\end{equation}
with $\beta = 8$, with $r_c$ being the King core radius \cite[see ][for a derivation of the equation]{1976ApJ...204...73I}. As suggested in \cite{2007ApJ...670..313S}, the King core radius can be related to the measurable Plummer radius by $r_c  = 0.64 r_{\mathrm{Plummer}}$. We use the tabulated list of Plummer radii in Table~6 of \cite{2007ApJ...670..313S}. The correlation strength of IMF and dynamical mass is similar and has a similar p-value as that with $\sigma_v$. As mentioned above, further analysis is ongoing to improve on the velocity dispersion estimates, and the simple relation of Eq.~\ref{eq:mtot} may not be best suited for all the 6 galaxies in our sample, making our estimates of the dynamical mass quite uncertain.
In the future, more carefully derived estimates will give the ability to confirm or dispute the presently observed (albeit weak) correlation of IMF slope with mass.

\paragraph{Ellipticity, half-light radii and absolute magnitudes}
These values are taken from \cite{2008ApJ...684.1075M}. No correlation between the IMF slope and any of these parameters is observed.

\subsection{Binary Fraction, Slope and $m_c$}

\begin{figure*}[!t]
\includegraphics[width=0.985\textwidth]{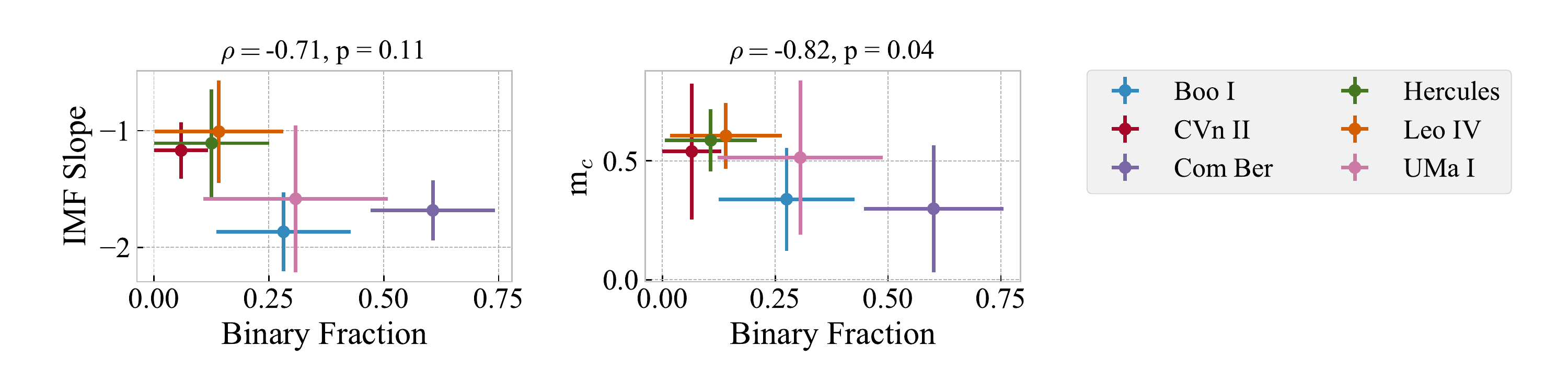}
\caption{Left: correlation between IMF slope and Binary Fraction. Right: correlation between $m_c$, the characteristic mass for a Log-normal model, and binary fraction. $\rho$ is the Pearson correlation coefficient, and the p-values represents the probability of obtaining an equivalent or larger values of $\rho$ from a sample with the same number of data but no correlation \label{fig:slpmcbf}.}
\end{figure*}

There appears to be a correlation between the best fit IMF slope values and binary fraction (see Fig.~\ref{fig:slpmcbf}). A similar correlation is observed between $m_c$ and the binary fraction for the Log-normal model. The fact that both slope and $m_c$ show this correlation with the binary fraction is a consequence of the fact that the slopes and $m_c$ are themselves correlated. Steeper slopes correspond to more bottom-heavy distributions, with lower mean mass, thus lower $m_c$.

This correlation with binary fraction is striking, and we investigated whether part of it could be spurious in origin. The individual triangle plots of Appendix~\ref{sec:corner_plots} show a weak correlation for the (slope, Binary Fraction) or ($m_c$, Binary Fraction) parameter pairs of the individual MCMC runs. 
Computing the Pearson correlation coefficient for the MCMC draws of the individual galaxies, we find $\rho = (-0.09,-0.25,0.05,-0.20,-0.12,-0.23)$ for the (slope, Binary Fraction) pairs and $\rho = (-0.10,0.01, 0.10, 0.02,\allowbreak-0.03,-0.11)$ for the ($m_c$, Binary Fraction) pairs, where the numbers correspond in order to (Boo~I, CVn~II, Com Ber, Hercules, Leo~IV, UMa~I).
We note that the correlation between value pairs across the UFDs sample (Fig.~\ref{fig:slpmcbf}) is much stronger than each of the individual correlations in the MCMC solution.

Given their weakness, these intrinsic correlations in the individual posterior distributions may only be partly responsible for the behavior observed in Fig.~\ref{fig:slpmcbf}. 
To corroborate this, we show in the bottom panel of Fig.~\ref{fig:ER} the Slope and $m_c$ marginal distributions when considering only MCMC draws around a fixed binary fraction value: $0.15 < \mathrm{Binary \, Fraction} < 0.25$. This range is appropriate for comparison with the system mass function of \cite{2003PASP..115..763C}, which assumes a binary fraction of 0.20 for the Milky Way.
Figure~\ref{fig:ER} shows that even when limited to this binary fraction range, our results for individual galaxies do not change significantly, and neither do the results over the UFDs sample. The fact that the results do not change when fixing the binary fraction further demonstrates that the observed correlation of Slope and $m_c$ with binary fraction over the UFDs cannot be solely explained by the intrinsic, small correlation of the two paramter pairs within the individual UFDs MCMC draws.

Still, it is hard to understand why some of the UFDs would ``prefer'' a shallower slope and smaller binary fraction while others prefer the opposite. 
We do not try to over-interpret the results for the binary fractions themselves. The binary fraction is clearly poorly constrained due to the small number statistics and all uncertainties related to e.g., the assumed mass ratio distribution. Leaving the binary fraction free to vary allows us to marginalize the pdf for the slope (or equivalently $m_c$) over this unknown parameter. Thus, rather than fixing the binary fraction to, e.g., the typical Milky Way value, we include it in our calculation to account for the uncertainty induced by its unknown value.

\section{Summary and Conclusions}
We have derived the parameters for a single power law and a log-normal model of the initial mass function of 6 ultra-faint dwarf satellites of the Milky Way. 
Our results show that, in the probed mass range of 0.5--0.8 M$_{\odot}$, the stellar populations of these objects have a mass distribution that differs from that of the Milky Way disk and young clusters stars, i.e. population I stars with solar metallicity.
When using a single power law model description for the IMF, the UFDs show a flatter slope than the -2.3 for the Galactic disk and open clusters \citep{1955ApJ...121..161S,2001MNRAS.322..231K,2012ApJ...748...14D,2017A&A...602A..22A}. When 
parametrizing the IMF as a Log-normal, the UFDs show a larger characteristic mass than the galactic value of 0.22-0.25 M$_{\odot}$ \citep{2003PASP..115..763C,2010AJ....139.2679B}. In either case the data show us that, in the 0.5-0.8 M$_{\odot}$ interval, there are more low-mass stars in the Milky Way relatively to the UFDs.
The significance of the discrepancy with the Milky Way IMF is stronger for the single power law model. The two-parameters Log-normal functional form is more flexible, thus constraints on the individual parameters, characteristic mass and width, are necessarily less stringent, and we cannot rule out the galactic values at more than $1\sigma$ for all of the UFDs. Given the limited mass range probed we do not attempt a quantitative model comparison between these two adopted parametrizations.

We investigate the possible environmental effects that may be driving this behavior and find that the average galactic metallicity moderately correlates with the IMF slope, while a lower, less significant correlation is observed between IMF slope and the velocity dispersion and the total galaxy mass. 

Even though all UFDs have more bottom light IMFs than the Milky Way disk, we observe variations within the UFDs sample.
The farthest UFDs, CVn~II, Hercules and Leo~IV have more bottom light IMFs than the Milky Way with respect to the UFDs closest to it, Boo~I, Com~Ber, UMa~I.
This could hint to some physical process that depends on the distance from the Milky Way as the responsible for the observed differences. However we caution that a higher fraction of undetected, residual fore- and background contamination may be affecting the closest UFDs, which have greater extent on the sky, and required more $HST$/ACS tiles to be observed. If that is the case, the less contaminated catalogs for the furthest UFDs would be showing a more authentic picture of these small satellites, with a much increased significance in IMF variations with respect to the Galaxy.

Further work is needed to try and increase the sample of studied systems, e.g., by using the James Webb Space Telescope and the Wide-Field Infrared Space Telescope, thus improving the reliability of the results suggested by our small sample.

\acknowledgements
We would like to thank Prof. Don Vandenberg (University of Victoria, British Columbia, Canada) 
for providing the oxygen-enhanced models used in this work, as well as to the \texttt{Fortran} 
routines used for their interpolation.

Support for program GO-12549 was provided by NASA through a grant from
the Space Telescope Science Institute, which is operated by the
Association of Universities for Research in Astronomy, Inc., under
NASA contract NAS 5-26555.  

This work was supported by a NASA Keck PI
Data Award, administered by the NASA Exoplanet Science Institute under
RSA number 1474359.  Data presented herein were obtained at the
W.M. Keck Observatory from telescope time allocated to NASA through
the agency's scientific partnership with the California Institute of
Technology and the University of California. The Observatory was made
possible by the generous financial support of the W.M. Keck
Foundation.  The authors wish to recognize and acknowledge the very
significant cultural role and reverence that the summit of Mauna Kea
has always had within the indigenous Hawaiian community. We are most
fortunate to have the opportunity to conduct observations from this
mountain.

ENK acknowledges support from NSF grant AST-1614081.

\appendix
\section{Validating The Fitting Technique}
\label{sec:validation}

In order to validate our method, we utilize simulated catalogs on which we run the algorithm described in Section~\ref{sec:method}. 
For the simulations, we adopt the star formation history and metallicity distribution function of Hercules. We simulated populations with both an underlying single power law IMF and a Log-normal one. All of the simulations have an intensity such that the number of observed stars is about 2000, similar to the numbers for our galaxies. All simulations assume a binary fraction of 30\%. For the single power law case, the adopted slope is -1.5, while for the Log-normal case we adopted the \cite{2003PASP..115..763C} values for the system IMF, ($m_c, \sigma) = (0.22, 0.57)$.

\begin{figure*}[ht!]
\gridline{\fig{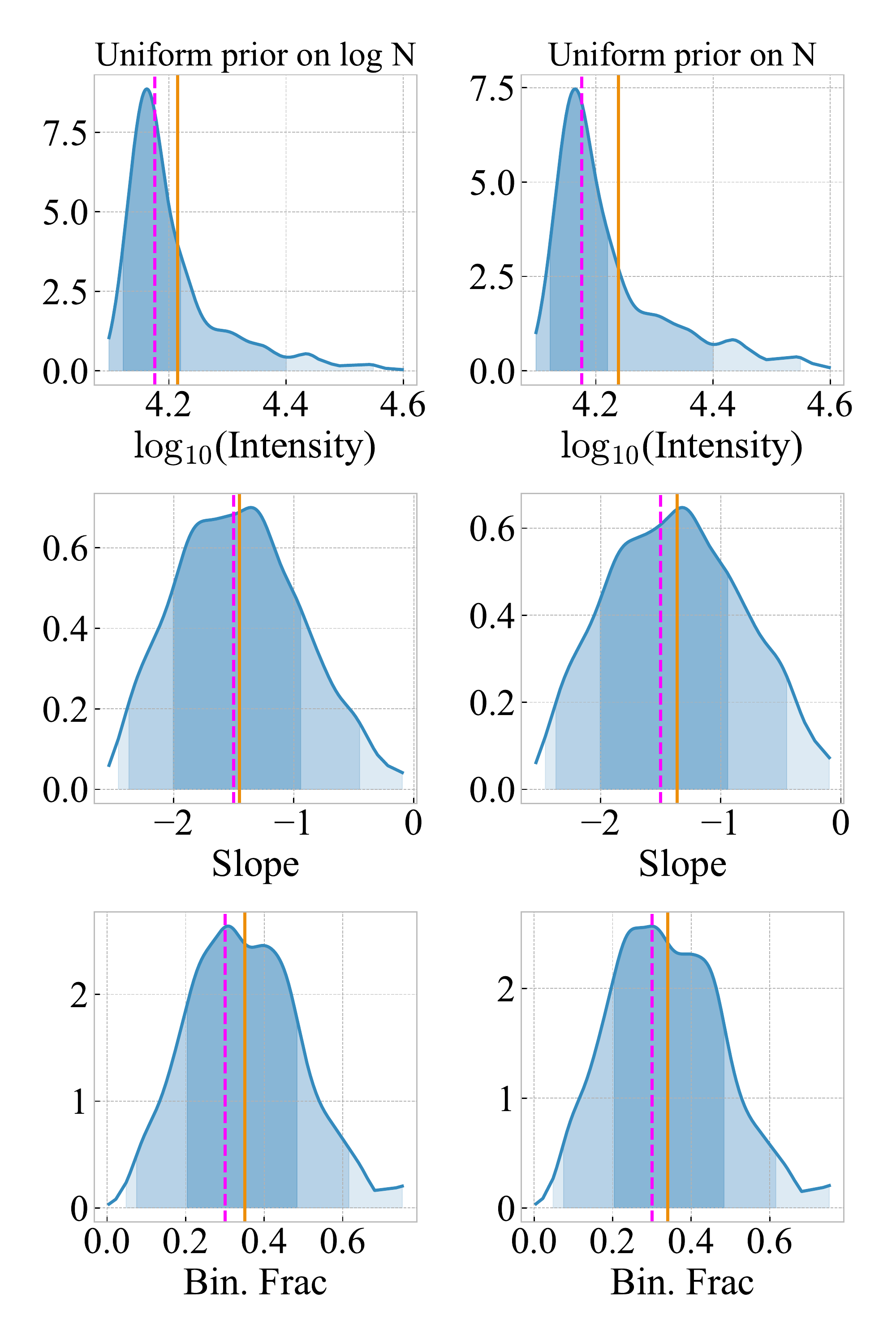}{0.315\textwidth}{Single Power Law (shallow)}
		 \fig{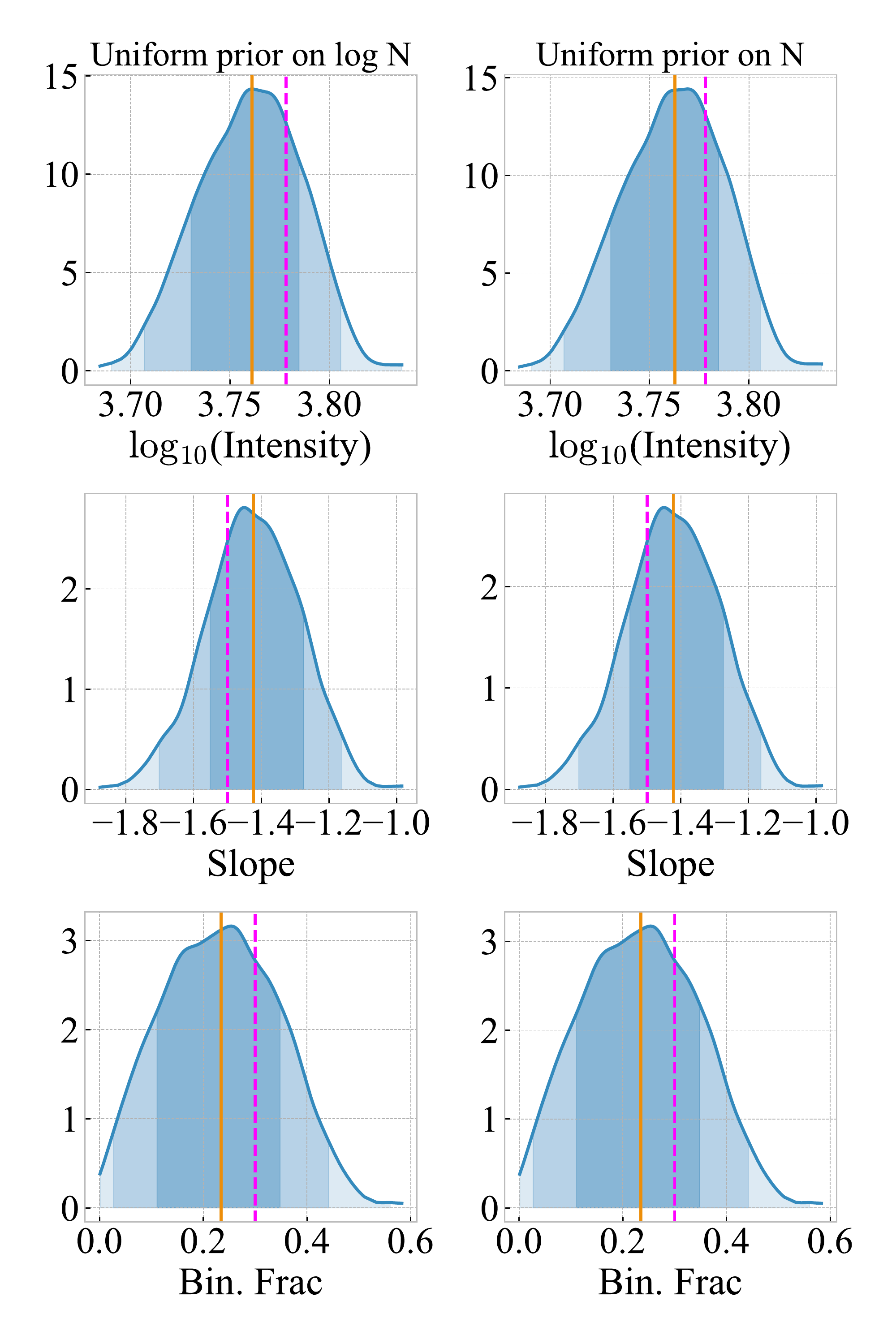}{0.315\textwidth}{Single Power law (deep)}
         \fig{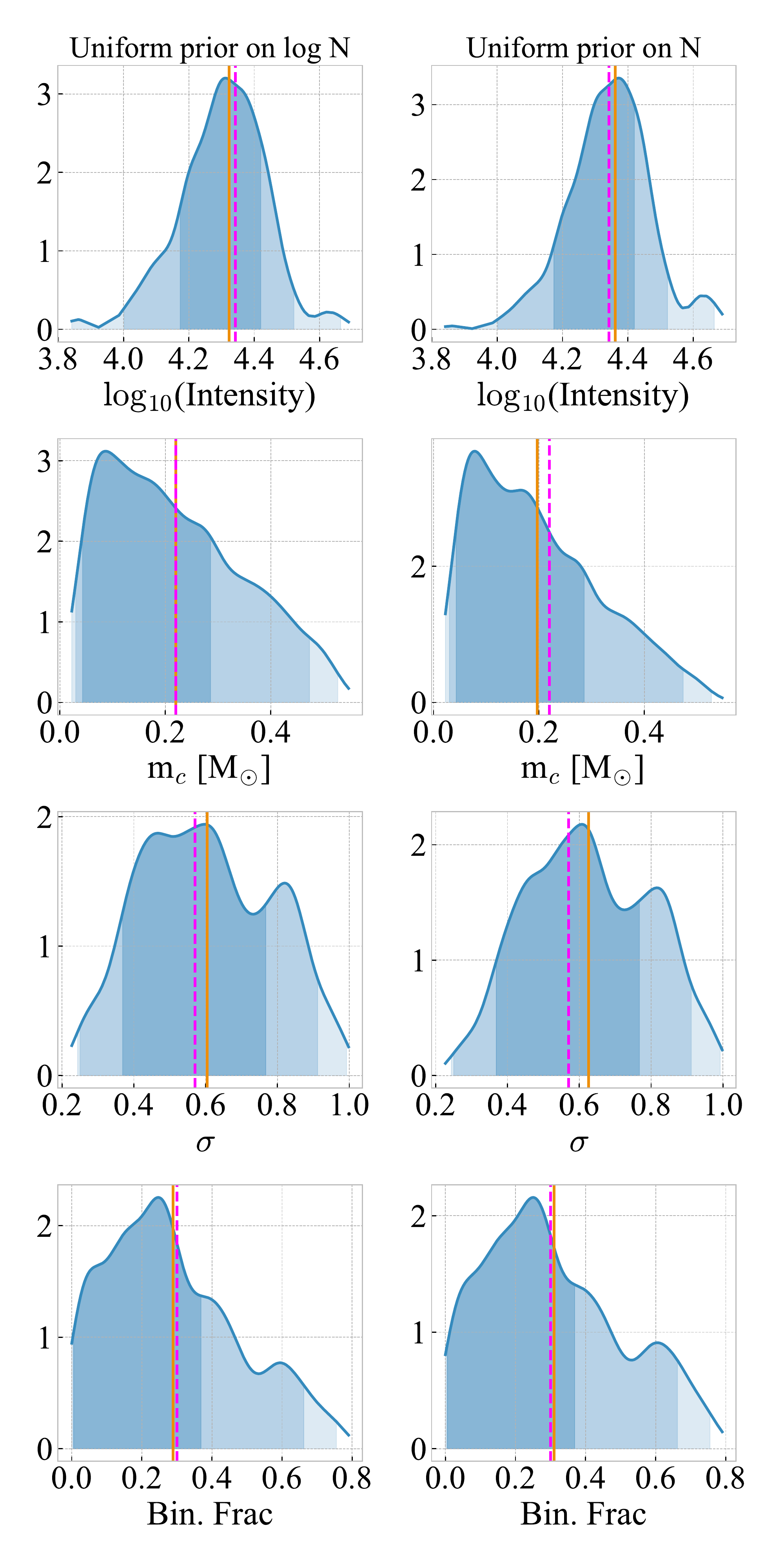}{0.31\textwidth}{Log Normal (shallow)}}
\caption{Results of a subset of the tests performed to validate the method. The curves represent kernel density estimates of the marginal distribution of the MCMC draws. The kernel width, $w$, is estimated using Scott's rule \citep{2015mdet.book.....S}: $w = \sigma  n^{-1./(d+4)}$, with $\sigma$ being the data variance, $n$ the number of points, $d$ the number of dimensions (always 1 in this case). The true values are indicated by the dashed magenta vertical lines, the best-fit values by the solid orange ones. The shaded areas indicate credibility intervals, estimated as the smallest intervals containing a $(0.6827, 0.9545, 0.9973)$ fraction of the probability. Left: single power law IMF case, for simulations of depths similar to our data. Center: single power law case, but for observations that are 3 magnitudes deeper; note the shrinking of the uncertainty intervals with respect to the left panel. Right: log-normal IMF case, again for simulations of depths similar to our data. Note that the m$_c$ value, for the logarithmic prior on the Intensity is recovered so precisely that the magenta and orange lines overlap. Also note that even though the peak value of 0.22 M$_{\odot}$ is outside the observed range (our simulations, like the observation reach about 0.5 M$_{\odot}$), it can be easily recovered within the uncertainties. \label{fig:sim_results}}
\end{figure*}

Figure~\ref{fig:sim_results} shows the results for some of these simulations. The left and right panels show results for simulated data of depths comparable to the observations of Hercules. The central panel shows results from an MCMC run on a simulated catalog where observations have been artificially made 3 magnitudes deeper, reaching about 0.15 $M_{\odot}$. The results in this case are obviously more precise than those for shallower observations.

We note that the estimates for the IMF parameters are always unbiased and the true values are easily recovered within the 68\% credible intervals for all cases and all variables. We are confident that, provided that the IMF model specification (single power law or log-normal) is the correct description of the underlying IMF for the ultra-faint dwarf populations, our method gives robust estimates for the model parameters.

In the Single Power Law (shallow) case the estimate of the best-fit value of the Intensity falls outside the 68\% credible interval. This is related to the fact that the marginal distributions are very skewed, with a long tail at high intensities. The mean, which we use to estimate the best-fit value, falls far from the mode. The credibility intervals instead typically straddle the mode, because for these pdfs, which are generally unimodal, the smallest interval containing a certain fraction of the probability (our definition of credible interval) does contain the mode region. We note that for the distributions obtained using the actual observations, this problem does not present itself, even though sometimes the mean is close to one edge of the 68\% credible interval.
This apparent problem arises from the fact that describing a complex multi-dimensional pdf in terms of a single central tendency value per dimension and some estimate of its uncertainty is a less than ideal process, because it can hardly capture the full available information.
We have chosen to use the mean as central tendency indicator, with the understanding that the uncertainty in our estimates can only be captured by using the full pdfs, which we always display.

We also note that the typical uncertainty from these tests is similar to the uncertainties on the fitted parameters for our real galaxies, thus our estimated uncertainties are reasonable.

\clearpage
\section{Detailed MCMC Results Plots}
\label{sec:corner_plots}

\begin{figure*}[ht!]
\gridline{\fig{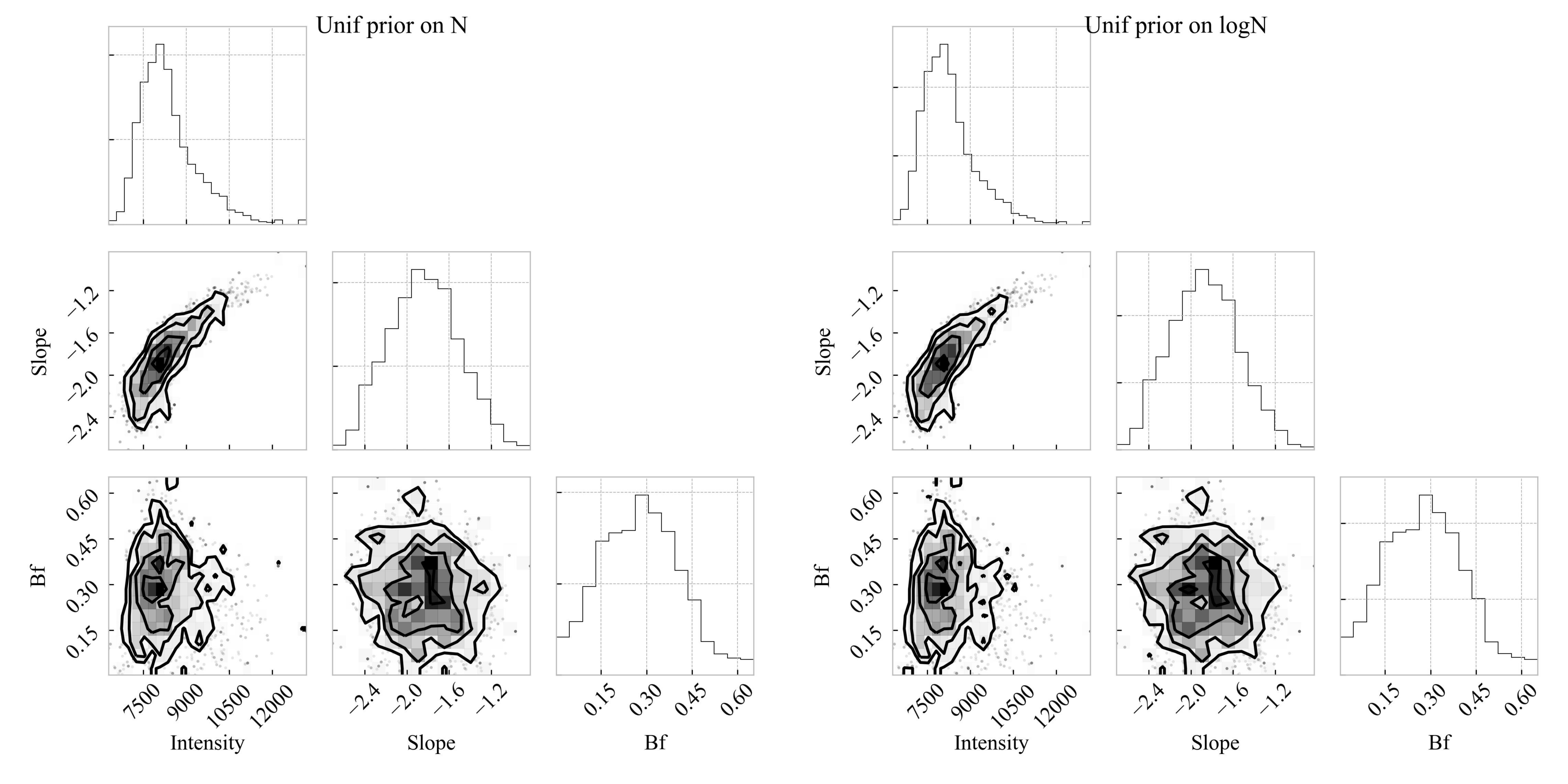}{0.9\textwidth}{Single Power Law model}}
\gridline{\fig{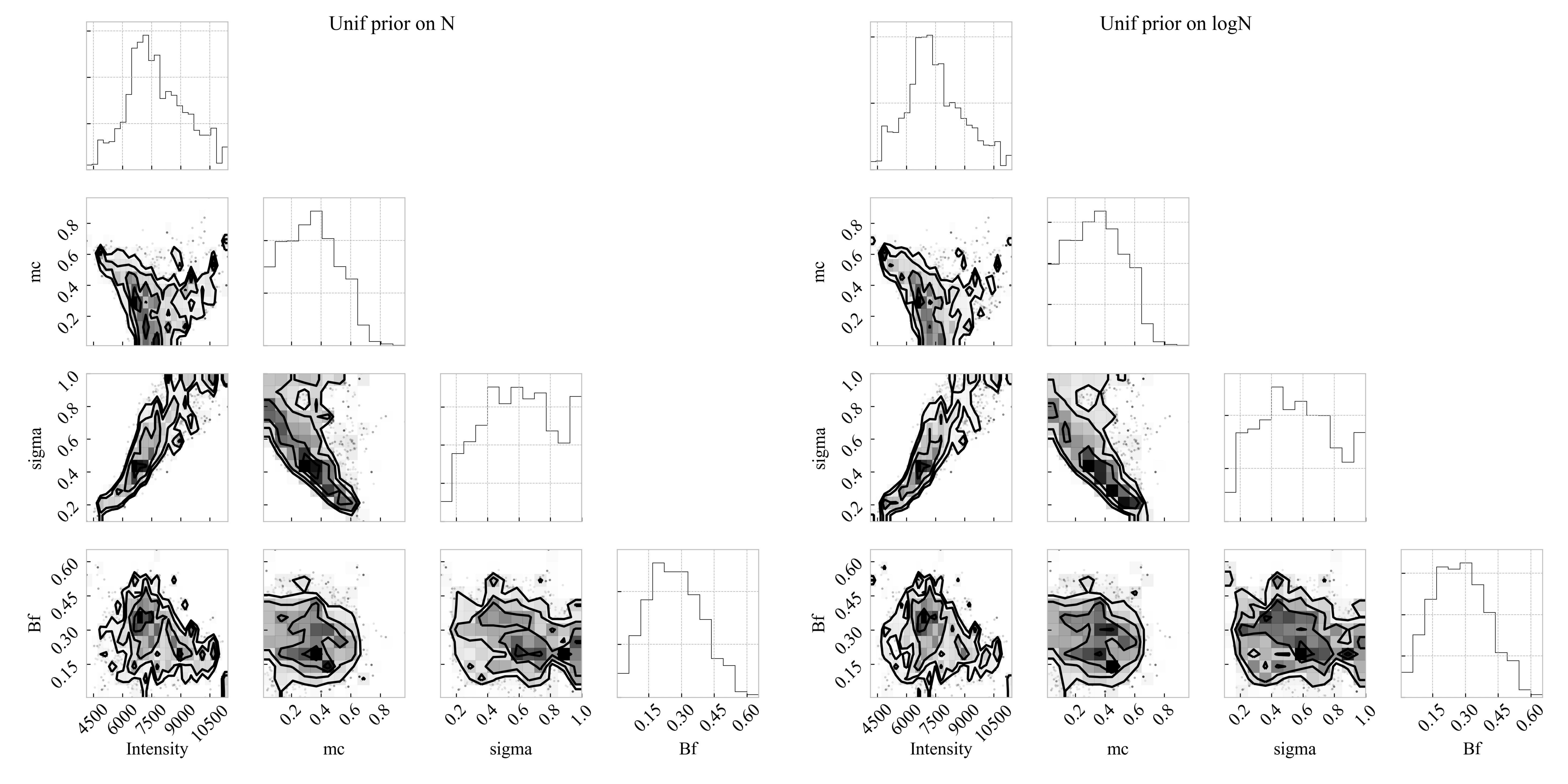}{0.9\textwidth}{Log-normal model}}
\caption{Samples from the posterior probability distribution function for the IMF model parameters, as given by our MCMC fitting technique. The case of the Boo~I galaxy.}
\end{figure*}

\begin{figure*}[ht!]
\gridline{\fig{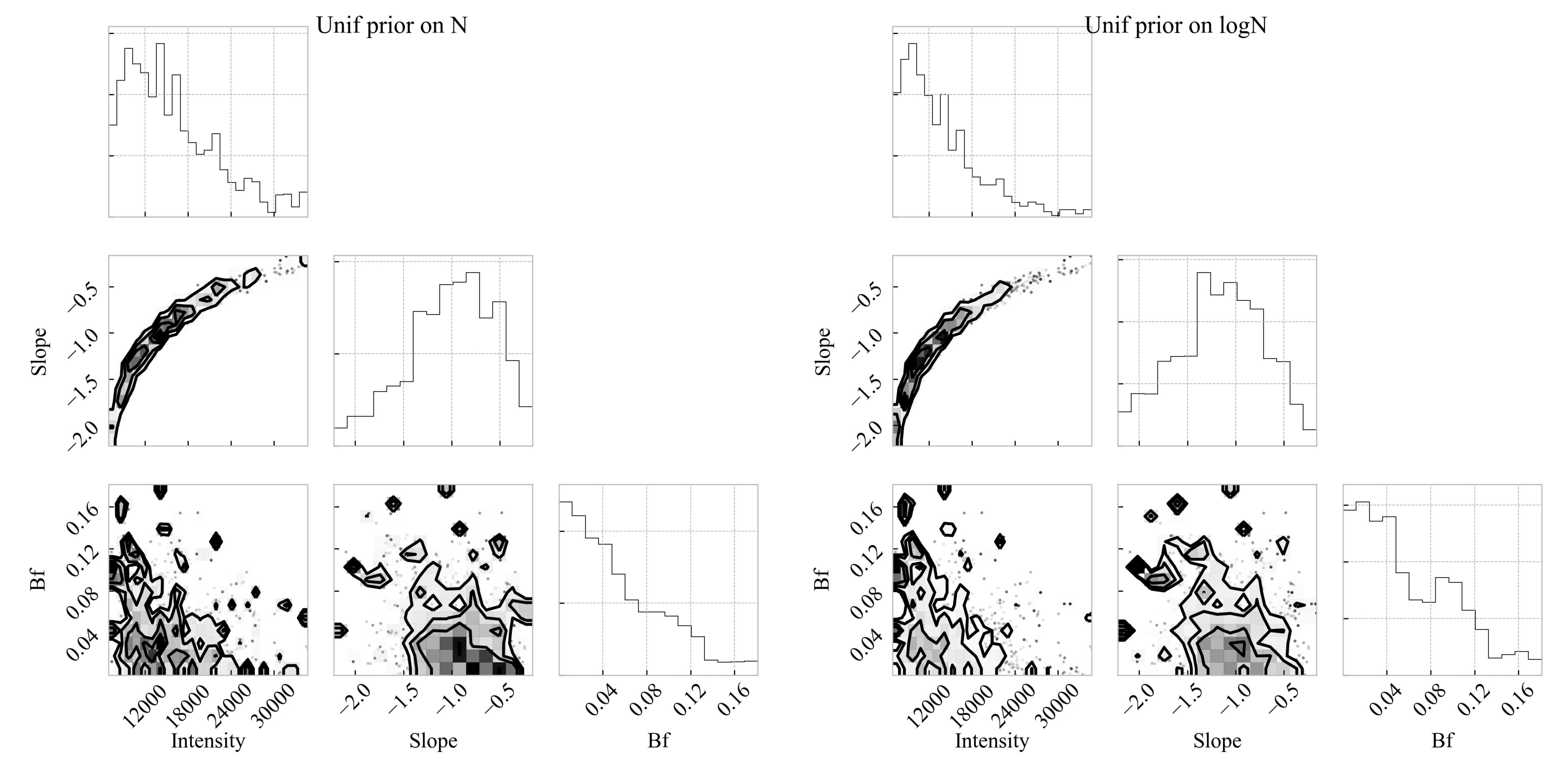}{0.9\textwidth}{Single Power Law model}}
\gridline{\fig{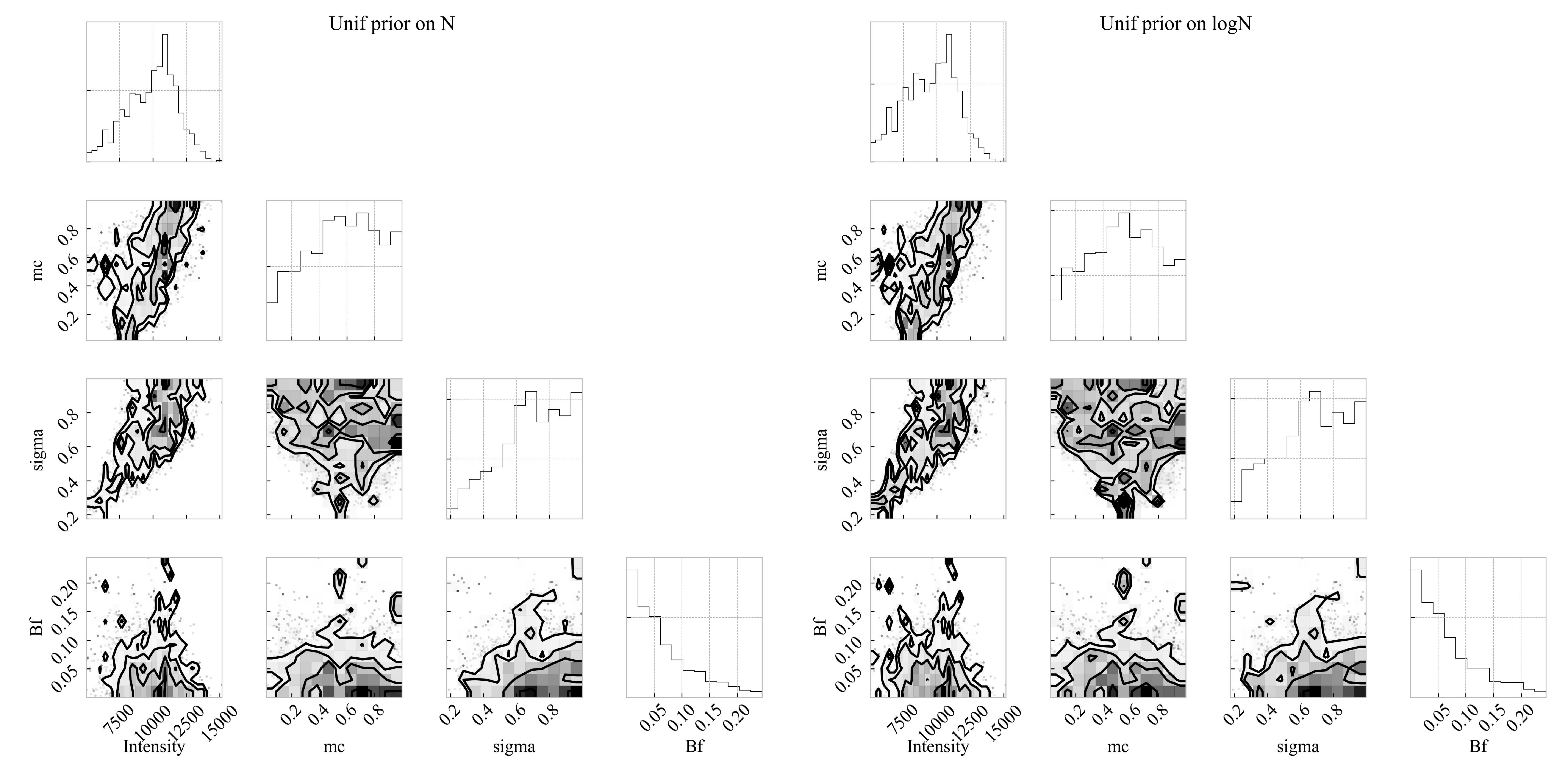}{0.9\textwidth}{Log-normal model}}
\caption{Samples from the posterior probability distribution function for the IMF model parameters, as given by our MCMC fitting technique. The case of the CVn~II galaxy.}
\end{figure*}

\begin{figure*}[ht!]
\gridline{\fig{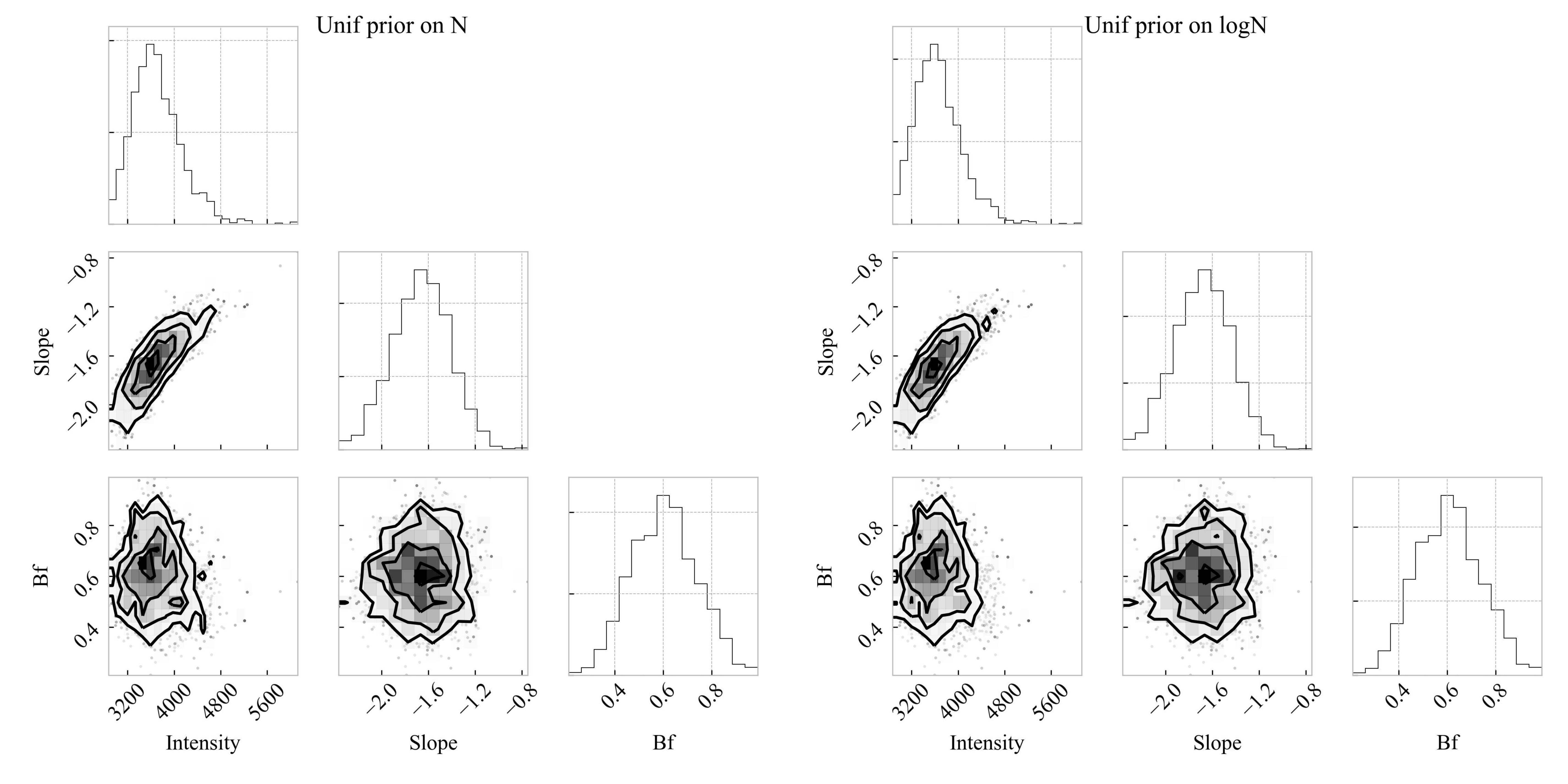}{0.9\textwidth}{Single Power Law model}}
\gridline{\fig{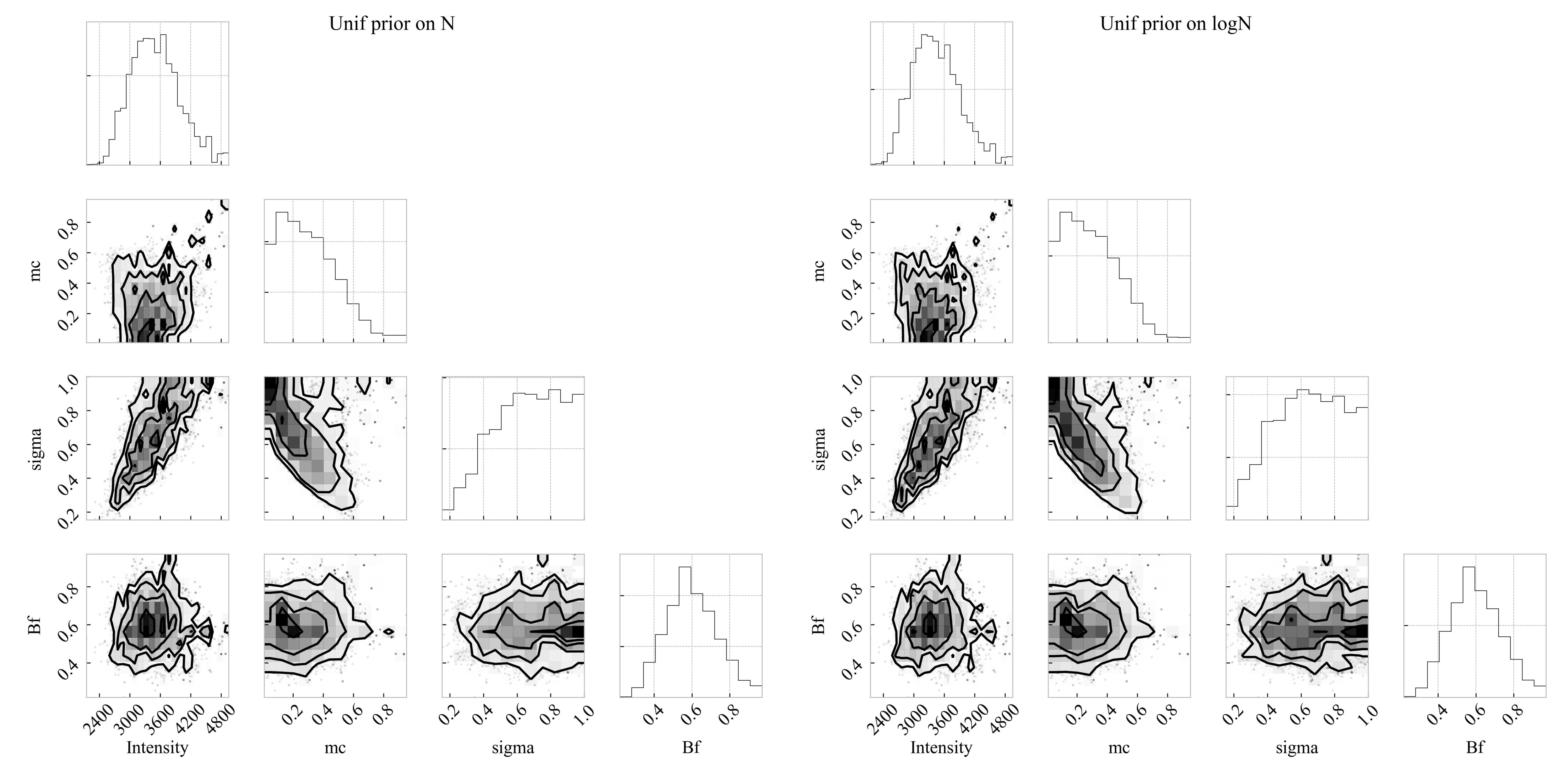}{0.9\textwidth}{Log-normal model}}
\caption{Samples from the posterior probability distribution function for the IMF model parameters, as given by our MCMC fitting technique. The case of the Com~Ber galaxy.}
\end{figure*}

\begin{figure*}[ht!]
\gridline{\fig{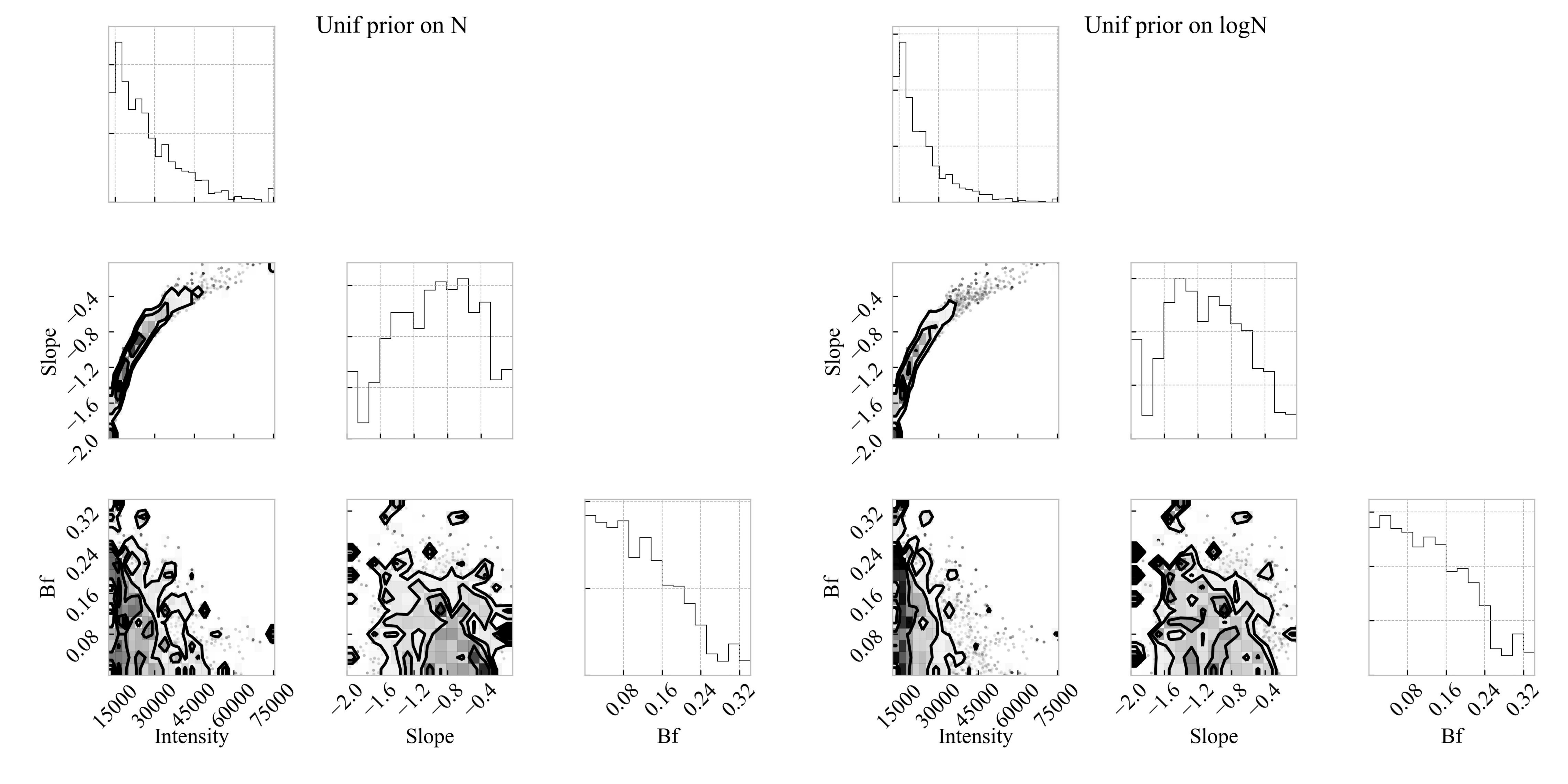}{0.9\textwidth}{Single Power Law model}}
\gridline{\fig{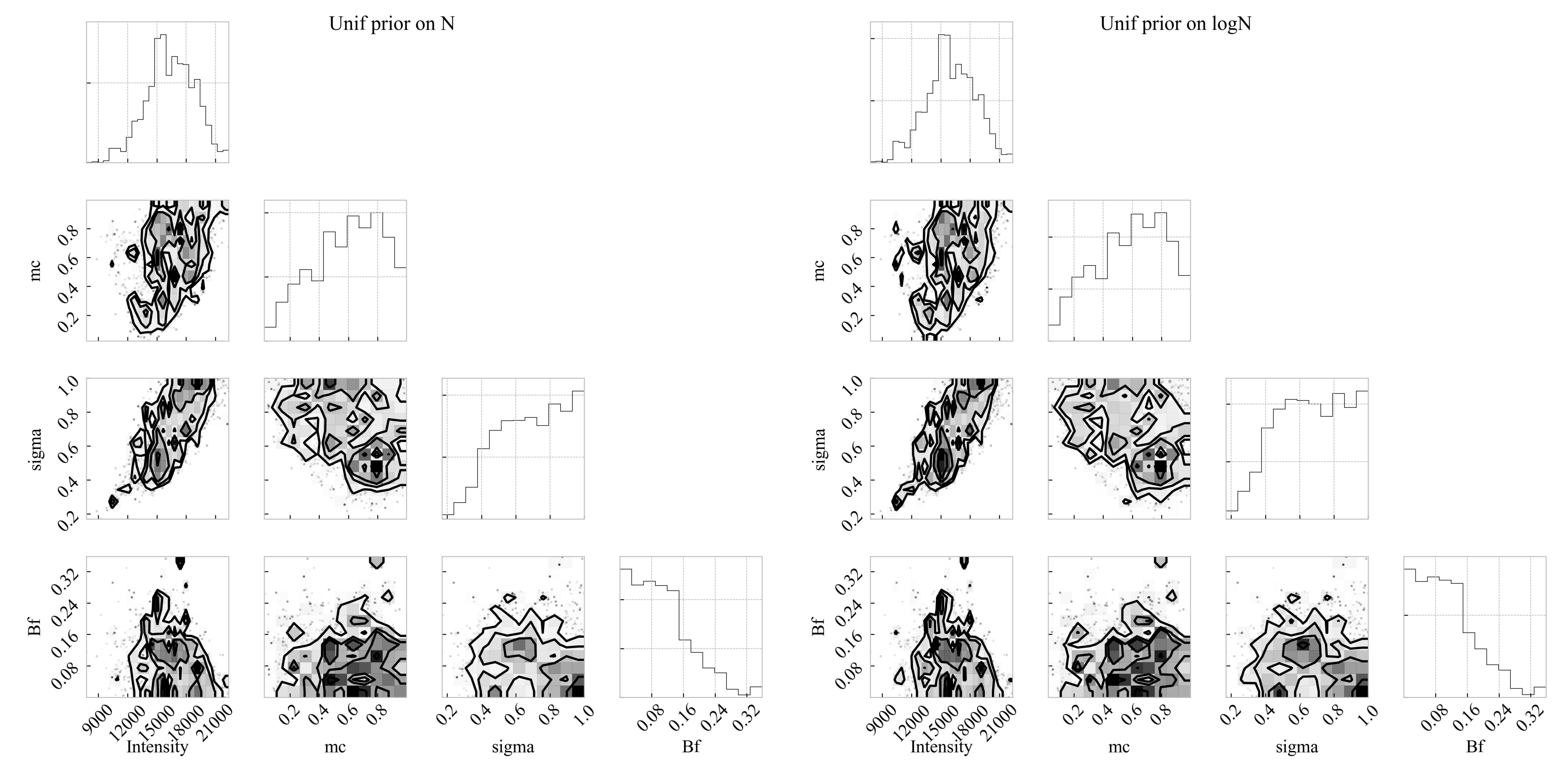}{0.9\textwidth}{Log-normal model}}
\caption{Samples from the posterior probability distribution function for the IMF model parameters, as given by our MCMC fitting technique. The case of the Hercules galaxy.}
\end{figure*}

\begin{figure*}[ht!]
\gridline{\fig{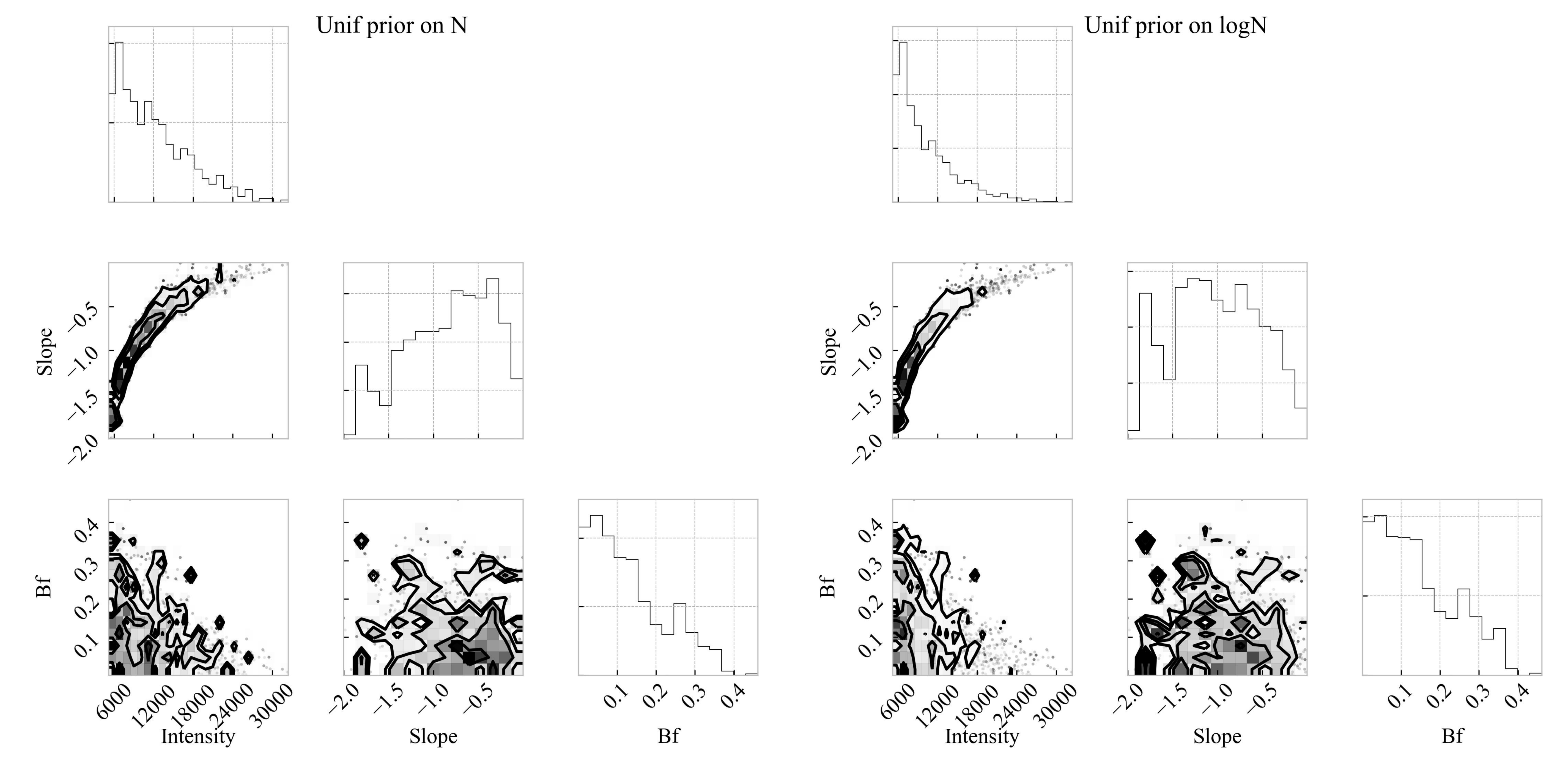}{0.9\textwidth}{Single Power Law model}}
\gridline{\fig{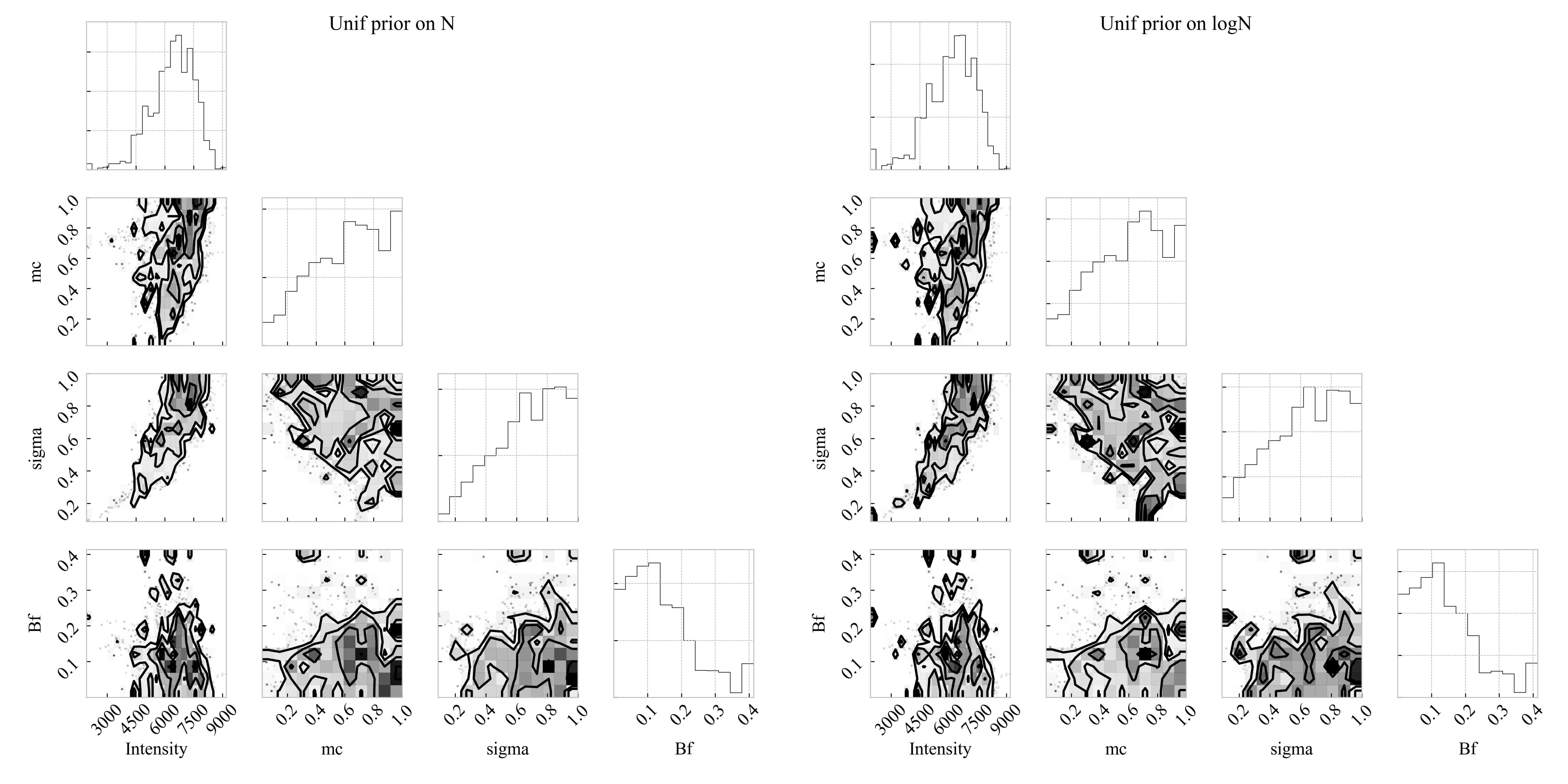}{0.9\textwidth}{Log-normal model}}
\caption{Samples from the posterior probability distribution function for the IMF model parameters, as given by our MCMC fitting technique. The case of the Leo~IV galaxy.}
\end{figure*}

\begin{figure*}[ht!]
\gridline{\fig{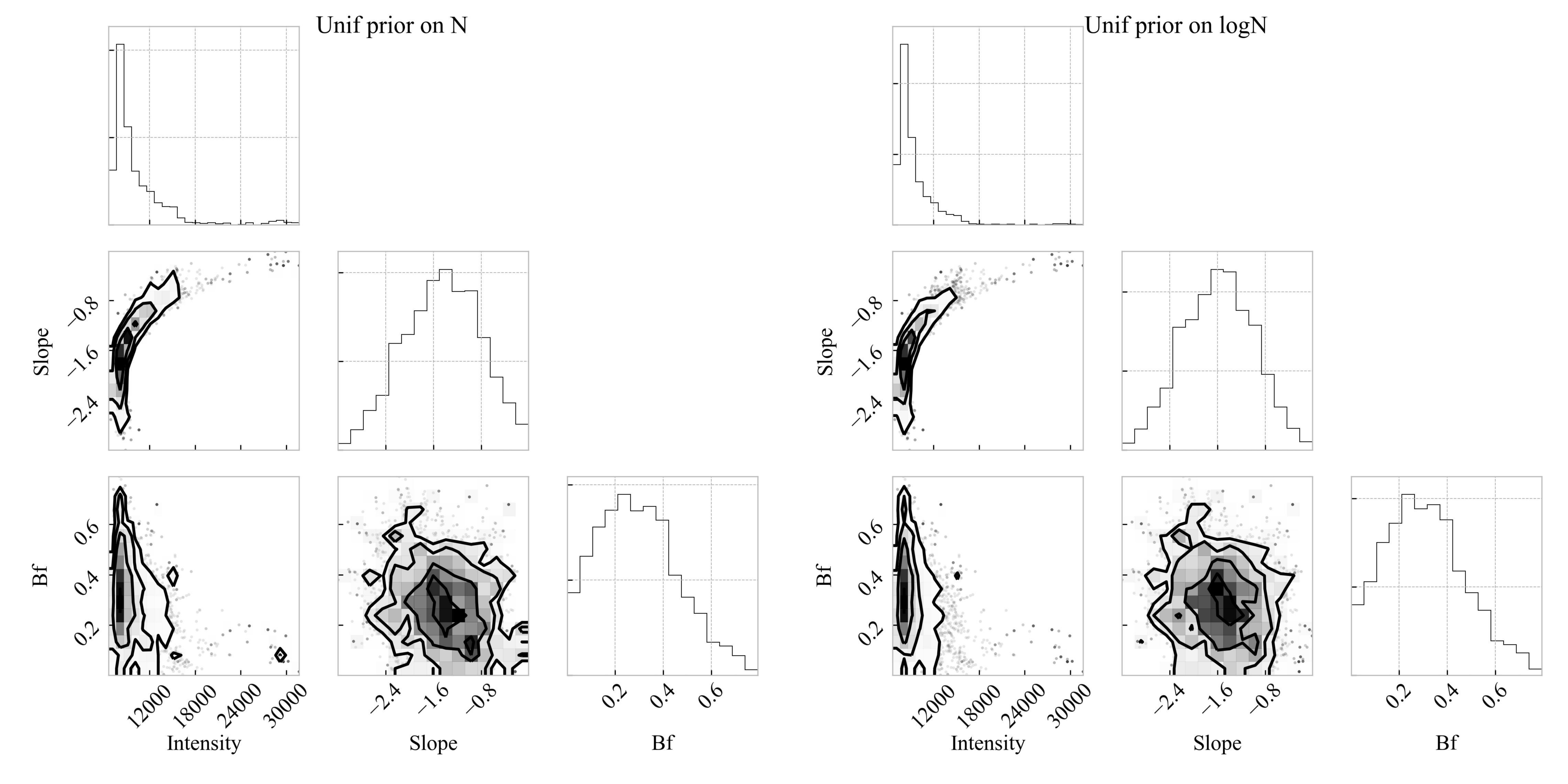}{0.9\textwidth}{Single Power Law model}}
\gridline{\fig{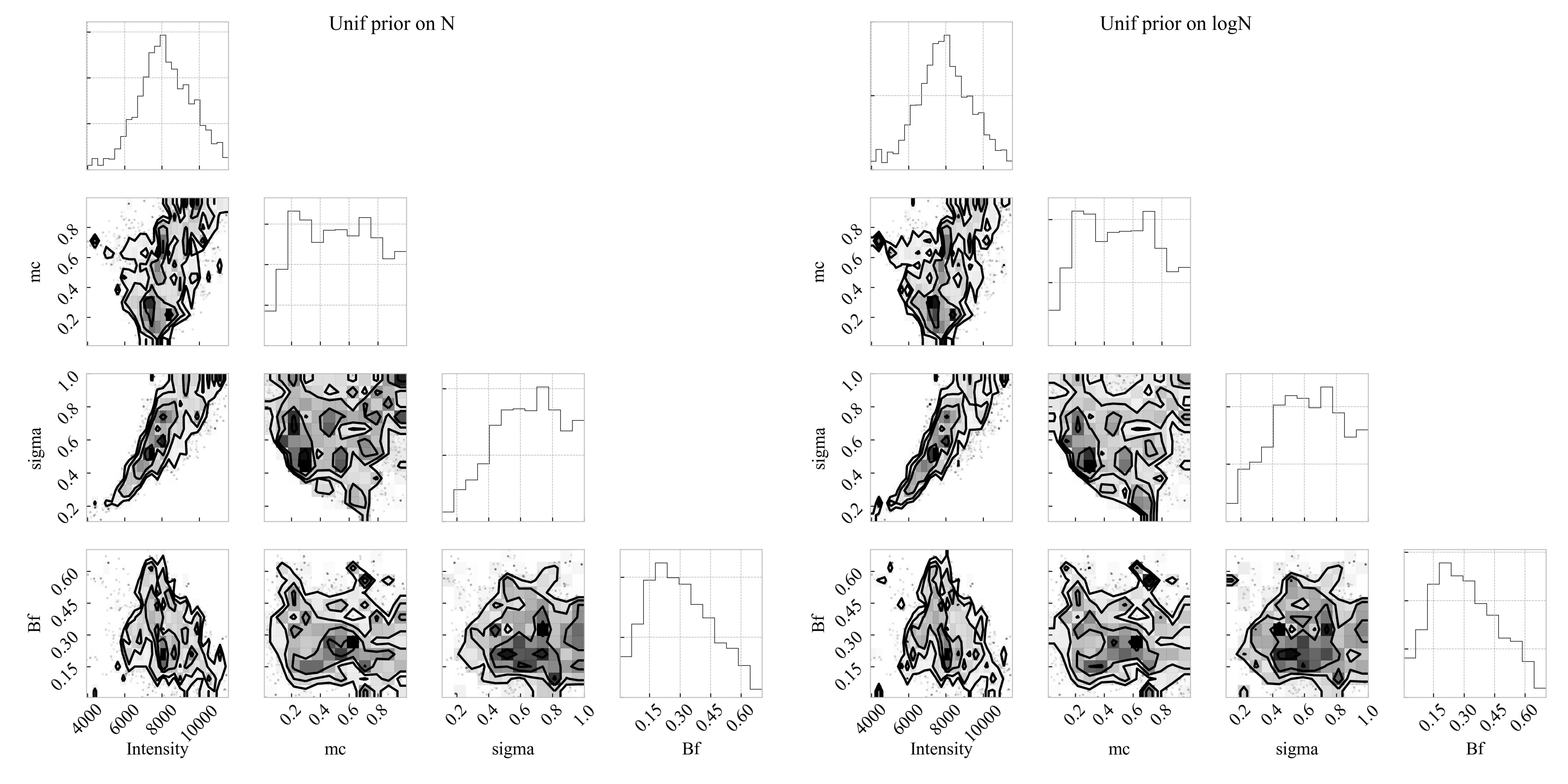}{0.9\textwidth}{Log-normal model}}
\caption{Samples from the posterior probability distribution function for the IMF model parameters, as given by our MCMC fitting technique. The case of the UMa~I galaxy.}
\end{figure*}

\clearpage
\section{Full Tables Of Results, Including Credible Intervals}
\label{sec:fulltabs}

\begin{deluxetable*}{lc||cc|cc||cc|cc||cc|cc}[h]
\tablecaption{Results for the single power law IMF model\label{tab:res_long_SPL}. Uniform and logarithmic column labels stand for parameter estimations with a uniform prior in Intensity or a prior that is uniform in $\log(\mathrm{Intensity})$, respectively. The credible intervals are defined as containing $(0.6827, 0.9545, 0.9973)$ of the total number of draws from the posterior (weighted by Intensity$^{-1}$ in the case of the logarithmic prior).}
\tablehead{
Galaxy & \colhead{Interval} &
\multicolumn{4}{c}{Intensity}  & 
\multicolumn{4}{c}{Slope}  & 
\multicolumn{4}{c}{Binary Fraction}  \\
\multicolumn{2}{c}{} &
\multicolumn{2}{c}{uniform} &
\multicolumn{2}{c}{logarithmic} &
\multicolumn{2}{c}{uniform} &
\multicolumn{2}{c}{logarithmic} &
\multicolumn{2}{c}{uniform} &
\multicolumn{2}{c}{logarithmic}
}
\tabletypesize{\footnotesize}
\startdata
         & Avg.      & 8.39e+3 &          &  8.28e+3 &          &  -1.84  &       &  -1.87  &       &  0.28  &      &  0.28  &      \\ 
         & 68\% & 7.15e+3 & 8.81e+3 &  7.12e+3 & 8.68e+3 &  -2.19  & -1.50 &  -2.20  & -1.53 &  0.14  & 0.40 &  0.14  & 0.40 \\ 
   Boo I & 95\% & 6.77e+3 & 1.05e+4 &  6.81e+3 & 1.03e+4 &  -2.44  & -1.20 &  -2.44  & -1.21 &  0.02  & 0.52 &  0.02  & 0.52 \\ 
         & 99\% & 6.52e+3 & 1.32e+4 &  6.30e+3 & 1.22e+4 &  -2.59  & -0.83 &  -2.71  & -0.97 &  0.01  & 0.65 &  0.01  & 0.65 \\ 
\hline
         & Avg.      & 1.58e+4 &          &  1.36e+4 &          &  -1.00  &       &  -1.17  &       &  0.05  &      &  0.06  &      \\ 
         & 68\% & 7.63e+3 & 1.82e+4 &  6.93e+3 & 1.49e+4 &  -1.31  & -0.43 &  -1.41  & -0.45 &  0.00  & 0.07 &  0.00  & 0.08 \\ 
  CVn II & 95\% & 6.93e+3 & 2.89e+4 &  6.93e+3 & 2.53e+4 &  -2.02  & -0.32 &  -2.02  & -0.36 &  0.00  & 0.14 &  0.00  & 0.15 \\ 
         & 99\% & 6.93e+3 & 3.46e+4 &  6.93e+3 & 3.46e+4 &  -2.22  & -0.16 &  -2.22  & -0.16 &  0.00  & 0.18 &  0.00  & 0.18 \\ 
\hline
         & Avg.      & 3.71e+3 &          &  3.67e+3 &          &  -1.66  &       &  -1.68  &       &  0.61  &      &  0.61  &      \\ 
         & 68\% & 3.24e+3 & 4.01e+3 &  3.19e+3 & 3.94e+3 &  -1.89  & -1.40 &  -1.94  & -1.45 &  0.47  & 0.75 &  0.47  & 0.75 \\ 
 Com Ber & 95\% & 2.93e+3 & 4.52e+3 &  2.92e+3 & 4.45e+3 &  -2.14  & -1.18 &  -2.15  & -1.19 &  0.36  & 0.88 &  0.37  & 0.88 \\ 
         & 99\% & 2.88e+3 & 5.25e+3 &  2.88e+3 & 5.25e+3 &  -2.37  & -1.06 &  -2.37  & -1.06 &  0.28  & 0.98 &  0.28  & 0.99 \\ 
\hline
         & Avg.      & 2.75e+4 &          &  2.32e+4 &          &  -0.93  &       &  -1.11  &       &  0.12  &      &  0.13  &      \\ 
         & 68\% & 1.37e+4 & 3.05e+4 &  1.37e+4 & 2.50e+4 &  -1.36  & -0.33 &  -1.57  & -0.61 &  0.00  & 0.15 &  0.00  & 0.16 \\ 
Hercules & 95\% & 1.26e+4 & 5.43e+4 &  1.26e+4 & 4.44e+4 &  -1.80  & -0.02 &  -2.00  & -0.32 &  0.00  & 0.27 &  0.00  & 0.29 \\ 
         & 99\% & 1.31e+4 & 7.56e+4 &  1.26e+4 & 7.56e+4 &  -2.00  & -0.02 &  -2.00  & -0.02 &  0.00  & 0.34 &  0.00  & 0.34 \\ 
\hline
         & Avg.      & 1.20e+4 &          &  1.01e+4 &          &  -0.82  &       &  -1.01  &       &  0.13  &      &  0.14  &      \\ 
         & 68\% & 5.28e+3 & 1.38e+4 &  5.22e+3 & 1.12e+4 &  -1.15  & -0.17 &  -1.45  & -0.40 &  0.00  & 0.16 &  0.00  & 0.18 \\ 
  Leo IV & 95\% & 5.28e+3 & 2.29e+4 &  5.18e+3 & 1.98e+4 &  -1.83  & -0.13 &  -1.82  & -0.20 &  0.00  & 0.31 &  0.01  & 0.34 \\ 
         & 99\% & 5.18e+3 & 2.95e+4 &  5.18e+3 & 2.85e+4 &  -1.86  & -0.00 &  -2.00  & -0.01 &  0.00  & 0.38 &  0.00  & 0.38 \\ 
\hline
         & Avg.      & 1.05e+4 &          &  9.56e+3 &          &  -1.46  &       &  -1.58  &       &  0.30  &      &  0.31  &      \\ 
         & 68\% & 7.17e+3 & 1.07e+4 &  7.16e+3 & 9.72e+3 &  -2.15  & -0.82 &  -2.21  & -1.00 &  0.09  & 0.43 &  0.11  & 0.44 \\ 
   UMa I & 95\% & 6.75e+3 & 1.86e+4 &  6.79e+3 & 1.50e+4 &  -2.71  & -0.22 &  -2.75  & -0.44 &  0.00  & 0.61 &  0.00  & 0.62 \\ 
         & 99\% & 6.82e+3 & 3.17e+4 &  6.63e+3 & 2.99e+4 &  -3.02  & -0.01 &  -3.03  & -0.06 &  0.00  & 0.74 &  0.00  & 0.74 \\ 
\hline
\hline
\enddata
\end{deluxetable*}

\begin{deluxetable*}{lc||cc|cc||cc|cc||cc|cc||cc|cc}
\tablecaption{Results for the log-normal IMF model\label{tab:res_long_LN}. Uniform and logarithmic column labels stand for parameter estimations with a uniform prior in Intensity or a prior that is uniform in $\log(\mathrm{Intensity})$, respectively. The credible intervals are defined as containing $(0.6827, 0.9545, 0.9973)$ of the total number of draws from the posterior (weighted by Intensity$^{-1}$ in the case of the logarithmic prior).}
\tablehead{
Galaxy & \colhead{Interval} &
\multicolumn{4}{c}{Intensity}  & 
\multicolumn{4}{c}{$m_c$}  & 
\multicolumn{4}{c}{$\sigma$}  & 
\multicolumn{4}{c}{Binary Fraction}  \\
\multicolumn{2}{c}{} &
\multicolumn{2}{c}{uniform} &
\multicolumn{2}{c}{logarithmic} &
\multicolumn{2}{c}{uniform} &
\multicolumn{2}{c}{logarithmic} &
\multicolumn{2}{c}{uniform} &
\multicolumn{2}{c}{logarithmic} &
\multicolumn{2}{c}{uniform} &
\multicolumn{2}{c}{logarithmic}
}
\tabletypesize{\footnotesize}
\startdata
        & Avg.      & 7.75e+3 &          &  7.46e+3 &          &  0.33 &      & 0.34 &      &      & 0.59 &      & 0.56 &      & 0.27 &      & 0.28 \\ 
         & 68\% & 6.40e+3 & 9.39e+3 &  5.78e+3 & 8.71e+3 &  0.13 & 0.51 & 0.12 & 0.52 & 0.29 & 0.80 & 0.29 & 0.80 & 0.13 & 0.38 & 0.13 & 0.38 \\ 
   Boo I & 95\% & 4.89e+3 & 1.07e+4 &  4.69e+3 & 1.05e+4 &  0.05 & 0.66 & 0.04 & 0.65 & 0.21 & 0.99 & 0.17 & 0.97 & 0.03 & 0.50 & 0.04 & 0.52 \\ 
         & 99\% & 4.13e+3 & 1.14e+4 &  4.13e+3 & 1.14e+4 &  0.01 & 0.77 & 0.01 & 0.76 & 0.10 & 0.99 & 0.10 & 0.99 & 0.01 & 0.65 & 0.01 & 0.65 \\ 
\hline
         & Avg.      & 9.91e+3 &          &  9.51e+3 &          &  0.56 &      & 0.54 &      &      & 0.70 &      & 0.67 &      & 0.06 &      & 0.07 \\ 
         & 68\% & 8.32e+3 & 1.21e+4 &  7.34e+3 & 1.14e+4 &  0.42 & 0.98 & 0.25 & 0.82 & 0.58 & 0.96 & 0.58 & 1.00 & 0.00 & 0.08 & 0.00 & 0.08 \\ 
  CVn II & 95\% & 6.04e+3 & 1.33e+4 &  5.55e+3 & 1.29e+4 &  0.11 & 0.99 & 0.10 & 0.99 & 0.32 & 1.00 & 0.28 & 0.99 & 0.00 & 0.19 & 0.00 & 0.19 \\ 
         & 99\% & 5.04e+3 & 1.42e+4 &  5.02e+3 & 1.40e+4 &  0.03 & 1.00 & 0.02 & 1.00 & 0.22 & 1.00 & 0.20 & 1.00 & 0.00 & 0.24 & 0.00 & 0.24 \\ 
\hline
         & Avg.      & 3.52e+3 &          &  3.45e+3 &          &  0.30 &      & 0.30 &      &      & 0.67 &      & 0.65 &      & 0.60 &      & 0.60 \\ 
         & 68\% & 2.99e+3 & 3.93e+3 &  2.92e+3 & 3.84e+3 &  0.04 & 0.41 & 0.03 & 0.39 & 0.55 & 0.98 & 0.49 & 0.94 & 0.44 & 0.71 & 0.45 & 0.72 \\ 
 Com Ber & 95\% & 2.63e+3 & 4.52e+3 &  2.56e+3 & 4.38e+3 &  0.01 & 0.66 & 0.02 & 0.65 & 0.31 & 1.00 & 0.29 & 1.00 & 0.35 & 0.88 & 0.35 & 0.88 \\ 
         & 99\% & 2.51e+3 & 4.95e+3 &  2.43e+3 & 4.95e+3 &  0.01 & 0.95 & 0.01 & 0.95 & 0.18 & 1.00 & 0.18 & 1.00 & 0.30 & 0.97 & 0.30 & 0.97 \\ 
\hline
         & Avg.      & 1.64e+4 &          &  1.60e+4 &          &  0.60 &      & 0.59 &      &      & 0.69 &      & 0.67 &      & 0.11 &      & 0.11 \\ 
         & 68\% & 1.43e+4 & 1.91e+4 &  1.41e+4 & 1.91e+4 &  0.46 & 0.94 & 0.46 & 0.95 & 0.58 & 1.00 & 0.49 & 0.94 & 0.00 & 0.13 & 0.01 & 0.14 \\ 
Hercules & 95\% & 1.16e+4 & 2.12e+4 &  1.04e+4 & 2.03e+4 &  0.16 & 0.99 & 0.15 & 0.98 & 0.36 & 1.00 & 0.34 & 1.00 & 0.00 & 0.25 & 0.00 & 0.25 \\ 
         & 99\% & 9.90e+3 & 2.23e+4 &  9.78e+3 & 2.23e+4 &  0.05 & 1.00 & 0.05 & 1.00 & 0.24 & 1.00 & 0.22 & 1.00 & 0.00 & 0.36 & 0.00 & 0.36 \\ 
\hline
         & Avg.      & 6.40e+3 &          &  6.14e+3 &          &  0.62 &      & 0.61 &      &      & 0.67 &      & 0.64 &      & 0.14 &      & 0.14 \\ 
         & 68\% & 5.72e+3 & 7.84e+3 &  5.18e+3 & 7.59e+3 &  0.48 & 0.99 & 0.47 & 0.98 & 0.59 & 1.00 & 0.51 & 0.98 & 0.02 & 0.18 & 0.02 & 0.19 \\ 
  Leo IV & 95\% & 4.38e+3 & 8.36e+3 &  3.77e+3 & 8.39e+3 &  0.17 & 1.00 & 0.15 & 0.99 & 0.27 & 1.00 & 0.22 & 1.00 & 0.00 & 0.34 & 0.00 & 0.35 \\ 
         & 99\%& 1.92e+3 & 8.51e+3 &  1.92e+3 & 8.48e+3 &  0.05 & 1.00 & 0.05 & 1.00 & 0.09 & 1.00 & 0.09 & 1.00 & 0.00 & 0.39 & 0.00 & 0.39 \\ 
\hline
         & Avg.      & 8.19e+3 &          &  7.92e+3 &          &  0.53 &      & 0.51 &      &      & 0.64 &      & 0.61 &      & 0.30 &      & 0.31 \\ 
         & 68\% & 6.88e+3 & 9.77e+3 &  6.19e+3 & 9.14e+3 &  0.20 & 0.77 & 0.19 & 0.74 & 0.49 & 0.94 & 0.40 & 0.88 & 0.08 & 0.40 & 0.12 & 0.45 \\ 
   UMa I & 95\% & 5.68e+3 & 1.11e+4 &  5.12e+3 & 1.10e+4 &  0.12 & 0.99 & 0.10 & 0.98 & 0.27 & 1.00 & 0.23 & 1.00 & 0.02 & 0.59 & 0.01 & 0.59 \\ 
         & 99\% & 4.14e+3 & 1.15e+4 &  4.02e+3 & 1.15e+4 &  0.03 & 1.00 & 0.02 & 1.00 & 0.13 & 1.00 & 0.12 & 1.00 & 0.00 & 0.67 & 0.00 & 0.67 \\ 
\hline\hline
\hline
\enddata
\end{deluxetable*}

\clearpage
\bibliography{biblio_UFDsIMF}

\begin{thebibliography}{}
\expandafter\ifx\csname natexlab\endcsname\relax\def\natexlab#1{#1}\fi
\providecommand{\url}[1]{\href{#1}{#1}}

\bibitem[{{Andersen} {et~al.}(2017){Andersen}, {Gennaro}, {Brandner}, {Stolte},
  {de Marchi}, {Meyer}, \& {Zinnecker}}]{2017A&A...602A..22A}
{Andersen}, M., {Gennaro}, M., {Brandner}, W., {et~al.} 2017, \aap, 602, A22

\bibitem[{{Aparicio} \& {Hidalgo}(2009)}]{Aparicio:2009uq}
{Aparicio}, A., \& {Hidalgo}, S.~L. 2009, \aj, 138, 558

\bibitem[{{Bastian} {et~al.}(2010){Bastian}, {Covey}, \&
  {Meyer}}]{2010ARA&A..48..339B}
{Bastian}, N., {Covey}, K.~R., \& {Meyer}, M.~R. 2010, \araa, 48, 339

\bibitem[{{Bernton} {et~al.}(2017){Bernton}, {Jacob}, {Gerber}, \&
  {Robert}}]{2017arXiv170105146B}
{Bernton}, E., {Jacob}, P.~E., {Gerber}, M., \& {Robert}, C.~P. 2017, ArXiv
  e-prints, arXiv:1701.05146

\bibitem[{{Bochanski} {et~al.}(2010){Bochanski}, {Hawley}, {Covey}, {West},
  {Reid}, {Golimowski}, \& {Ivezi{\'c}}}]{2010AJ....139.2679B}
{Bochanski}, J.~J., {Hawley}, S.~L., {Covey}, K.~R., {et~al.} 2010, \aj, 139,
  2679

\bibitem[{{Brown} {et~al.}(2014){Brown}, {Tumlinson}, {Geha}, {Simon},
  {Vargas}, {VandenBerg}, {Kirby}, {Kalirai}, {Avila}, {Gennaro}, {Ferguson},
  {Mu{\~n}oz}, {Guhathakurta}, \& {Renzini}}]{2014ApJ...796...91B}
{Brown}, T.~M., {Tumlinson}, J., {Geha}, M., {et~al.} 2014, \apj, 796, 91

\bibitem[{{Cameron} \& {Pettitt}(2012)}]{2012MNRAS.425...44C}
{Cameron}, E., \& {Pettitt}, A.~N. 2012, \mnras, 425, 44

\bibitem[{{Cappellari} {et~al.}(2012){Cappellari}, {McDermid}, {Alatalo},
  {Blitz}, {Bois}, {Bournaud}, {Bureau}, {Crocker}, {Davies}, {Davis}, {de
  Zeeuw}, {Duc}, {Emsellem}, {Khochfar}, {Krajnovi{\'c}}, {Kuntschner},
  {Lablanche}, {Morganti}, {Naab}, {Oosterloo}, {Sarzi}, {Scott}, {Serra},
  {Weijmans}, \& {Young}}]{2012Natur.484..485C}
{Cappellari}, M., {McDermid}, R.~M., {Alatalo}, K., {et~al.} 2012, \nat, 484,
  485

\bibitem[{{Cardelli} {et~al.}(1989){Cardelli}, {Clayton}, \&
  {Mathis}}]{1989ApJ...345..245C}
{Cardelli}, J.~A., {Clayton}, G.~C., \& {Mathis}, J.~S. 1989, \apj, 345, 245

\bibitem[{{Chabrier}(2003)}]{2003PASP..115..763C}
{Chabrier}, G. 2003, \pasp, 115, 763

\bibitem[{{Cignoni} {et~al.}(2006){Cignoni}, {Degl'Innocenti}, {Prada Moroni},
  \& {Shore}}]{Cignoni:2006vn}
{Cignoni}, M., {Degl'Innocenti}, S., {Prada Moroni}, P.~G., \& {Shore}, S.~N.
  2006, \aap, 459, 783

\bibitem[{{Conroy} \& {van Dokkum}(2012)}]{2012ApJ...760...71C}
{Conroy}, C., \& {van Dokkum}, P.~G. 2012, \apj, 760, 71

\bibitem[{{Da Rio} {et~al.}(2012){Da Rio}, {Robberto}, {Hillenbrand},
  {Henning}, \& {Stassun}}]{2012ApJ...748...14D}
{Da Rio}, N., {Robberto}, M., {Hillenbrand}, L.~A., {Henning}, T., \&
  {Stassun}, K.~G. 2012, \apj, 748, 14

\bibitem[{{Dolphin}(2002)}]{Dolphin:2002lr}
{Dolphin}, A.~E. 2002, \mnras, 332, 91

\bibitem[{{Fitzpatrick}(1999)}]{1999PASP..111...63F}
{Fitzpatrick}, E.~L. 1999, \pasp, 111, 63

\bibitem[{Foreman-Mackey(2016)}]{corner}
Foreman-Mackey, D. 2016, The Journal of Open Source Software, 24,
  doi:10.21105/joss.00024.
\newblock \url{http://dx.doi.org/10.5281/zenodo.45906}

\bibitem[{{Foreman-Mackey} {et~al.}(2013){Foreman-Mackey}, {Hogg}, {Lang}, \&
  {Goodman}}]{2013PASP..125..306F}
{Foreman-Mackey}, D., {Hogg}, D.~W., {Lang}, D., \& {Goodman}, J. 2013, \pasp,
  125, 306

\bibitem[{{Geha} {et~al.}(2013){Geha}, {Brown}, {Tumlinson}, {Kalirai},
  {Simon}, {Kirby}, {VandenBerg}, {Mu{\~n}oz}, {Avila}, {Guhathakurta}, \&
  {Ferguson}}]{2013ApJ...771...29G}
{Geha}, M., {Brown}, T.~M., {Tumlinson}, J., {et~al.} 2013, \apj, 771, 29

\bibitem[{{Gennaro} {et~al.}(2015){Gennaro}, {Tchernyshyov}, {Brown}, \&
  {Gordon}}]{2015ApJ...808...45G}
{Gennaro}, M., {Tchernyshyov}, K., {Brown}, T.~M., \& {Gordon}, K.~D. 2015,
  \apj, 808, 45

\bibitem[{{Goodman} \& {Weare}(2010)}]{2010CAMCS...5...65G}
{Goodman}, J., \& {Weare}, J. 2010, Communications in Applied Mathematics and
  Computational Science, Vol.~5, No.~1, p.~65-80, 2010, 5, 65

\bibitem[{{Harris} \& {Zaritsky}(2001)}]{Harris:2001fj}
{Harris}, J., \& {Zaritsky}, D. 2001, \apjs, 136, 25

\bibitem[{{Illingworth}(1976)}]{1976ApJ...204...73I}
{Illingworth}, G. 1976, \apj, 204, 73

\bibitem[{{Kalirai} {et~al.}(2013){Kalirai}, {Anderson}, {Dotter}, {Richer},
  {Fahlman}, {Hansen}, {Hurley}, {Reid}, {Rich}, \&
  {Shara}}]{2013ApJ...763..110K}
{Kalirai}, J.~S., {Anderson}, J., {Dotter}, A., {et~al.} 2013, \apj, 763, 110

\bibitem[{{Kroupa}(2001)}]{2001MNRAS.322..231K}
{Kroupa}, P. 2001, \mnras, 322, 231

\bibitem[{{Kroupa}(2002)}]{2002Sci...295...82K}
---. 2002, Science, 295, 82

\bibitem[{{Leigh} {et~al.}(2012){Leigh}, {Umbreit}, {Sills}, {Knigge}, {de
  Marchi}, {Glebbeek}, \& {Sarajedini}}]{2012MNRAS.422.1592L}
{Leigh}, N., {Umbreit}, S., {Sills}, A., {et~al.} 2012, \mnras, 422, 1592

\bibitem[{{Li} {et~al.}(2017){Li}, {Simon}, {Drlica-Wagner}, {Bechtol}, {Wang},
  {Garc{\'{\i}}a-Bellido}, {Frieman}, {Marshall}, {James}, {Strigari}, {Pace},
  {Balbinot}, {Zhang}, {Abbott}, {Allam}, {Benoit-L{\'e}vy}, {Bernstein},
  {Bertin}, {Brooks}, {Burke}, {Carnero Rosell}, {Carrasco Kind}, {Carretero},
  {Cunha}, {D'Andrea}, {da Costa}, {DePoy}, {Desai}, {Diehl}, {Eifler},
  {Flaugher}, {Goldstein}, {Gruen}, {Gruendl}, {Gschwend}, {Gutierrez},
  {Krause}, {Kuehn}, {Lin}, {Maia}, {March}, {Menanteau}, {Miquel}, {Plazas},
  {Romer}, {Sanchez}, {Santiago}, {Schubnell}, {Sevilla-Noarbe}, {Smith},
  {Sobreira}, {Suchyta}, {Tarle}, {Thomas}, {Tucker}, {Walker}, {Wechsler},
  {Wester}, {Yanny}, \& {(DES Collaboration}}]{2017ApJ...838....8L}
{Li}, T.~S., {Simon}, J.~D., {Drlica-Wagner}, A., {et~al.} 2017, \apj, 838, 8

\bibitem[{{Marjoram} {et~al.}(2003){Marjoram}, {Molitor}, {Plagnol}, \&
  {Tavar{\'e}}}]{2003PNAS..10015324M}
{Marjoram}, P., {Molitor}, J., {Plagnol}, V., \& {Tavar{\'e}}, S. 2003,
  Proceedings of the National Academy of Science, 100, 15324

\bibitem[{{Martin} {et~al.}(2008){Martin}, {de Jong}, \&
  {Rix}}]{2008ApJ...684.1075M}
{Martin}, N.~F., {de Jong}, J.~T.~A., \& {Rix}, H.-W. 2008, \apj, 684, 1075

\bibitem[{{Newman} {et~al.}(2017){Newman}, {Smith}, {Conroy}, {Villaume}, \&
  {van Dokkum}}]{2017ApJ...845..157N}
{Newman}, A.~B., {Smith}, R.~J., {Conroy}, C., {Villaume}, A., \& {van Dokkum},
  P. 2017, \apj, 845, 157

\bibitem[{{Ng} {et~al.}(2002){Ng}, {Brogt}, {Chiosi}, \&
  {Bertelli}}]{Ng:2002gf}
{Ng}, Y.~K., {Brogt}, E., {Chiosi}, C., \& {Bertelli}, G. 2002, \aap, 392, 1129

\bibitem[{{Offner} {et~al.}(2014){Offner}, {Clark}, {Hennebelle}, {Bastian},
  {Bate}, {Hopkins}, {Moraux}, \& {Whitworth}}]{2014prpl.conf...53O}
{Offner}, S.~S.~R., {Clark}, P.~C., {Hennebelle}, P., {et~al.} 2014, Protostars
  and Planets VI, 53

\bibitem[{Phillips \&
  Venkatasubramanian(2011)}]{DBLP:journals/corr/abs-1103-1625}
Phillips, J.~M., \& Venkatasubramanian, S. 2011, CoRR, abs/1103.1625.
\newblock \url{http://arxiv.org/abs/1103.1625}

\bibitem[{Pritchard {et~al.}(1999)Pritchard, Seielstad, Perez-Lezaun, \&
  Feldman}]{Pritchard:1999td}
Pritchard, J.~K., Seielstad, M.~T., Perez-Lezaun, A., \& Feldman, M.~W. 1999,
  Mol. Biol. Evol., 16, 1791

\bibitem[{{Robin} {et~al.}(2003){Robin}, {Reyl{\'e}}, {Derri{\`e}re}, \&
  {Picaud}}]{2003A&A...409..523R}
{Robin}, A.~C., {Reyl{\'e}}, C., {Derri{\`e}re}, S., \& {Picaud}, S. 2003,
  \aap, 409, 523

\bibitem[{Rubin(1984)}]{rubin1984}
Rubin, D.~B. 1984, Ann. Statist., 12, 1151.
\newblock \url{https://doi.org/10.1214/aos/1176346785}

\bibitem[{{Salpeter}(1955)}]{1955ApJ...121..161S}
{Salpeter}, E.~E. 1955, \apj, 121, 161

\bibitem[{{Scott}(2015)}]{2015mdet.book.....S}
{Scott}, D.~W. 2015, {Multivariate Density Estimation: Theory, Practice, and
  Visualization}

\bibitem[{{Simon} \& {Geha}(2007)}]{2007ApJ...670..313S}
{Simon}, J.~D., \& {Geha}, M. 2007, \apj, 670, 313

\bibitem[{{Smith}(2014)}]{2014MNRAS.443L..69S}
{Smith}, R.~J. 2014, \mnras, 443, L69

\bibitem[{{Spitzer}(1987)}]{1987degc.book.....S}
{Spitzer}, L. 1987, {Dynamical evolution of globular clusters}

\bibitem[{{van Dokkum} \& {Conroy}(2010)}]{2010Natur.468..940V}
{van Dokkum}, P.~G., \& {Conroy}, C. 2010, \nat, 468, 940

\bibitem[{{van Dokkum} \& {Conroy}(2011)}]{2011ApJ...735L..13V}
---. 2011, \apjl, 735, L13

\bibitem[{{van Dokkum} \& {Conroy}(2012)}]{2012ApJ...760...70V}
---. 2012, \apj, 760, 70

\bibitem[{{VandenBerg} {et~al.}(2014){VandenBerg}, {Bergbusch}, {Ferguson}, \&
  {Edvardsson}}]{2014ApJ...794...72V}
{VandenBerg}, D.~A., {Bergbusch}, P.~A., {Ferguson}, J.~W., \& {Edvardsson}, B.
  2014, \apj, 794, 72

\bibitem[{{Vergely} {et~al.}(2002){Vergely}, {K{\"o}ppen}, {Egret}, \&
  {Bienaym{\'e}}}]{Vergely:2002ve}
{Vergely}, J.-L., {K{\"o}ppen}, J., {Egret}, D., \& {Bienaym{\'e}}, O. 2002,
  \aap, 390, 917

\bibitem[{{Webb} {et~al.}(2017){Webb}, {Vesperini}, {Dalessandro}, {Beccari},
  {Ferraro}, \& {Lanzoni}}]{2017MNRAS.471.3845W}
{Webb}, J.~J., {Vesperini}, E., {Dalessandro}, E., {et~al.} 2017, \mnras, 471,
  3845

\bibitem[{{Weyant} {et~al.}(2013){Weyant}, {Schafer}, \&
  {Wood-Vasey}}]{2013ApJ...764..116W}
{Weyant}, A., {Schafer}, C., \& {Wood-Vasey}, W.~M. 2013, \apj, 764, 116

\end{thebibliography}
\bibliographystyle{aasjournal}

\end{document}